\documentclass[twocolumn,showpacs,preprintnumbers,amsmath,amssymb]{revtex4}

\usepackage{graphicx}
\usepackage{dcolumn}
\usepackage{bm}

\begin{document}

\preprint{APS/123-QED}

\title{Quantum criticality around metal-insulator transitions
of 
strongly correlated electrons}

\author{Takahiro Misawa}
\email{misawa@solis.t.u-tokyo.ac.jp}
\author{Masatoshi Imada}%
\affiliation{%
Department of Applied Physics, University of Tokyo,
7-3-1 Hongo, Bunkyo-ku, Tokyo, 113-8656, Japan
}%

\date{\today}

\begin{abstract}
Quantum criticality of metal-insulator transitions in correlated electron systems
is shown to belong to an unconventional universality class with violation of 
Ginzburg-Landau-Wilson (GLW) scheme formulated for symmetry breaking transitions.  
This unconventionality arises from an emergent character of the quantum critical point, 
which appears at the marginal point between
the Ising-type symmetry breaking at nonzero temperatures and the topological transition of the 
Fermi surface at zero temperature. We show that Hartree-Fock approximations of 
an extended Hubbard model on square lattices
are capable of such metal-insulator transitions 
with  unusual criticality under a preexisting symmetry breaking.
The obtained universality is consistent with the scaling theory formulated for Mott transition 
and with a number of numerical results beyond the mean-field level,
implying that 
the preexisting symmetry breaking is not necessarily 
required for the emergence of this unconventional universality.
Examinations of fluctuation effects indicate that the obtained critical exponents 
remain essentially exact beyond the mean-field level. 
It further clarifies the whole structure of singularities by a unified treatment of 
the band-width-control and filling-control transitions.
Detailed analyses on the criticality, containing diverging carrier density fluctuations 
around the marginal quantum critical point, are presented from microscopic calculations and 
reveal the nature as quantum critical ``opalescence".  
The mechanism of emerging marginal quantum critical point is ascribed to a positive feedback and interplay 
between the preexisting gap formation present even in metals and kinetic energy gain (loss) of 
the metallic carrier.  
Analyses on crossovers between GLW type at nonzero temperature 
and topological type at zero temperature show that the critical exponents observed  
in (V,Cr)$_2$O$_3$ and $\kappa$-ET-type organic conductor provide us with evidences for the existence of 
the present marginal quantum criticality. 

\end{abstract}

\maketitle

\section{Introduction}
\label{sec:Introduction}
Metal insulator (MI) transitions in correlated electron systems have been a challenging 
subject of studies for decades~\cite{RMP}.
In this paper, we show from microscopic analyses that MI transitions of correlated electrons 
bear an unconventional feature in view of quantum phase transitions.  
This is shown by taking an example of the Hartree-Fock approximation of the Hubbard model 
in two dimensions. 
One might suspect whether the Hartree-Fock approximations could yield a phase transition which
is beyond the conventional scheme. However, we show that the simple 
mean-field approximation allows us to get insight into this issue by capturing the correct 
interplay of symmetry-breaking and topological characters of the transition.
We also discuss that the unconventionality survives beyond the limitation of the mean-field study, 
where the scaling theory~\cite{Imada_94,Imada_95} formulated for the Mott transition is applicable.
To make the motivation of the present study clearer, we describe somewhat detailed 
introduction to clarify why the present study has a significance and how the unconventionality arises  
in view of quantum phase transitions in general.  Then in the latter part of the introduction, we 
review the previous studies on MI transitions to position the starting point.

\subsection{General perspective in view of quantum phase transitions}
A ubiquitous example of phase transitions seen in nature is the transition between 
gas and liquid.
At the critical point of the gas-liquid phase transition, the compressibility 
\begin{equation}
\kappa=-\frac{1}{V}\frac{dV}{dP}=\frac{1}{n}\frac{dn}{dP}=\frac{1}{n^2}\frac{dn}{d\mu}
\label{kappa}
\end{equation}
and the equivalent density fluctuations $\int dr \langle n(r)n(0) \rangle$ diverge, where $V$ and $\mu$ are the volume and chemical potential, respectively.
The opalescence observed in light scattering is caused by such diverging density fluctuations. 
Nearly a century ago, the critical opalescence found in the scale of the light wave length was analyzed 
by Ornstein, Zernike~\cite{Ornstein}, Einstein and Smoluchovski~\cite{Einstein, Smoluchovski} as one of the evidences 
for the existence of atoms.  

The gas-liquid transition is mapped to the ferromagnetic transition in the Ising model
with the $Z_2$ symmetry expressed by the Hamiltonian
\begin{equation}
H=-J\sum_{\langle i,j\rangle }S_{i}S_{j} - h\sum_i S_i,
\label{Ising}
\end{equation}
for the local spin $S_i$ taking $\pm 1$ at the site $i$, and interacting with short-ranged neighbors at $j$
and under the symmetry breaking field $h$.
The Ising model represents the simplest example that phase transitions in general take place because of 
a spontaneous breaking of symmetries originally hold in the Hamiltonian at $h=0$. 
Although the gas-liquid transition at a first glance does not look breaking the symmetry 
as in the ordered phase of the Ising model,
the transition is actually mapped to the $Z_2$ symmetry breaking along the first-order coexistence boundary.
In fact, the up and down spin phases in the Ising model are mapped to the gas and liquid phases, respectively,
from which the both transitions turn out to belong to the same universality class.
In more general, thermal phase transitions are all characterized by a certain type of spontaneous symmetry breakings.

Phase transitions by the conventional spontaneous symmetry breaking is correctly described by the Ginzburg-Landau-Wilson
(GLW) scheme~\cite{GinzburgLandau,Wilson,Nigel}. 
In this scheme, the free energy $F$ is expressed by a functional of a spatially dependent order parameter $m(r)$ after eliminating other degrees of freedom.
$F$ is assumed to allow a regular expansion with respect to $m(r)$ such as
\begin{eqnarray}
	F&=&\int dr [ -h m(r)+ \frac{a}{2!}m(r)^2+ \frac{K}{2!}(\nabla m(r))^2 + \frac{b}{4!}m(r)^4 \nonumber \\
&&+ \cdots ],
\label{GLFreeEnergy}
\end{eqnarray}
where $b$ and $K$ are positive constants and the coefficient $a$ depends
on the control parameter $g$ around the critical point $g_{c}$
as $a=a_{0}(g-g_{c})$.

\begin{figure}[h!]
		\begin{center}
			\includegraphics[width=3cm,clip]{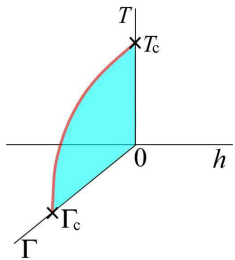}   
		\end{center}
	\caption{(color online)
	Schematic phase diagram of Ising model under longitudinal and transverse magnetic fields, 
	$h$ and $\Gamma$,
                 respectively. Up and down spin phases are separated
                 by the first-order boundary illustrated by shaded(blue)
                 sheet below the critical line shown by bold(brown) curve 
                 in parameter space of $T$, $h$ and the transverse Ising field $\Gamma$.
                 Beyond the quantum critical point at $\Gamma_c$, the transition disappears and the up and down spin
                 phases are adiabatically connected even through the route on the $T=0$ plane.}
	\label{QCPSchematic}
	\end{figure}

This classical picture of the phase transitions is modified in quantum systems.
For instance, when the transverse field $- \Gamma\sum_i S^x_i$ is additionally applied to the Ising model 
(\ref{Ising}) by considering spin-1/2 quantum spins ${\bf S}_i=(S_i^x,S_i^y,S_i^z)$, 
$T_c$ is suppressed to zero at a critical value $\Gamma_c$ and 
the transition disappears as illustrated in Fig.~\ref{QCPSchematic}.
This may be regarded as a destruction of the symmetry broken order by quantum fluctuations instead of 
thermal fluctuations.  

In fact, a quantum system may be expressed in the path integral formalism, where an additional dimension representing the 
imaginary time or inverse-temperature axis is introduced in addition to the $d$-dimensional spatial degrees of freedom. 
Although the added imaginary time direction could be qualitatively different from the spatial dimensions, 
one can show that it simply corresponds to $z$ additional spatial dimensions, where $z$ is called the dynamical exponent.
The transverse Ising model is simply represented by $z=1$. Namely, 
the quantum phase transition 
at $T=0$ in $d$ spatial dimensions realized by increasing $\Gamma$ to $\Gamma_c$ may be mapped to 
the transition at nonzero temperature in $d+1$ dimensional classical systems~\cite{Elliott}.

However, in more general, if the quantum dynamics has a coupling to dissipative gapless excitations 
such as particle-hole excitations in metals, $z$ may become larger than unity such as 
two or three~\cite{Moriya,Hertz,Millis}. Ferromagnetic and antiferromagnetic phase transitions in metals 
have been extensively studied from this viewpoint. 
The quantum critical point appears when the critical temperature is lowered to zero.
 Around the quantum critical point, low-energy excitations arising from growing magnetic correlations
have been proposed to cause non-Fermi liquid properties~\cite{Moriya,Hertz,Millis}. 
A particular type of metamagnetic transitions has also been
studied around the ``quantum critical end point"~\cite{QCEPMillis},
where an apparent symmetry does not change at the critical point,
whereas an implicit symmetry breaking does occur as in the case of the
gas-liquid transition. 
Even in these cases with $z$ larger than unity, however, 
the GLW scheme remains essentially valid when we consider a proper $d+z$-dimensional system.
  
Here the dynamical exponent $z$ is defined from the ratio of the temporal correlation length 
$\xi_t$ to the spatial one $\xi$
of the order parameter $m(r,t)$ in $d$ dimensions. 
In the case of conventional quantum phase transitions, $\xi_t$ and $\xi$ are defined by
\begin{align}
\xi=-\lim_{r \rightarrow \infty}\left(\frac{\log\langle m(r,0)m(0,0) \rangle}{r}\right)^{-1} \label{eq:con_xi}, \\
\xi_{t}=-\lim_{t \rightarrow \infty}\left(\frac{\log\langle m(0,t)m(0,0) \rangle}{t}\right)^{-1}.\label{eq:con_xi_t}
\end{align}
respectively.
More concretely, the dynamical exponent $z$
is defined in the asymptotic limit to the critical point $g\rightarrow g_c$ by
\begin{eqnarray}
z=\lim_{g\rightarrow g_c}  \frac{\log{\xi_{t}}}{\log {\xi}}. \label{eqn:Schi1}
\end{eqnarray}
As we will show in the next subsection, in the case of the metal-insulator
transitions, these definitions should be modified,
because the correlations in metals decay not exponentially but with power 
laws as a function of distance in general.

In contrast to the description of the phase transition by the symmetry breaking, several classes of quantum phase transitions at $T=0$ are not  
captured by such extensions of the GLW scheme and are not characterized by the spontaneous symmetry breaking.  
One of the simplest examples is found in the transition between a metal and a band insulator.
The ground-state energy $E$ of noninteracting electrons near the parabolic band bottom with the 
single-particle dispersion $\epsilon =Ak^2$ in $d$ dimensions is obviously described by 
\begin{equation}
E\propto n^{(d+2)/d} \propto \mu^{(d+2)/2},
\label{MIT}
\end{equation}
in the metallic phase for the electron density $n$ and the chemical potential $\mu$ measured from the 
band bottom. On the other hand, throughout the insulating phase with the chemical potential varied in the gap, $E=0$ holds
because of $n=0$.
This trivially nonanalytic form of the energy as a function of $\mu$ clearly represents
a quantum phase transition with the singularity at $\mu=0$, but it is equally obvious 
that it does not follow the GLW scheme.
This transition of noninteracting fermions is not accompanied by any kind of 
spontaneous symmetry breaking. 
It is rather characterized only by the presence or absence of the Fermi surface,
which is a topological difference.
Wen~\cite{Wen} has proposed a category of ``quantum orders" for this class of transitions.
  
\begin{figure}[h!]
		\begin{center}
			\includegraphics[width=3.5cm,clip]{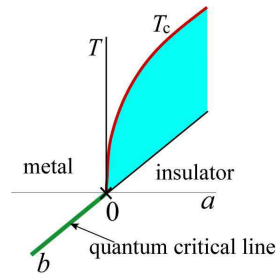}   
		\end{center}
	\caption{(color online) Proposed phase diagram of MI transition for strongly correlated 
                 electron systems. 
                A crucial difference from Fig.~\ref{QCPSchematic} is that the quantum critical line starts 
                   along the $b$ axis (the thick (green) line along $a=T=0$ ) beyond the endpoint
               of the first-order boundary called marginal quantum critical point depicted by the cross. 
               This difference is a direct consequence of the topological nature of MI transition at $T=0$.
                 Here the first-order boundary is again shown by the shaded (blue) surface. }
	\label{MQCPSchematic}
	\end{figure}

In this paper, we show that an involved interplay of the topological nature at $T=0$ with the conventional
GLW character at $T \ne 0$ emerges for metal-insulator (MI) transitions of correlated electrons.
By controlling a parameter for quantum fluctuations, a zero-temperature critical line characterized by 
the topological transition of Fermi surface from a metal to an insulator switches over to a finite-temperature
critical line characterized by GLW scheme.
The phase diagram is now seriously modified from Fig.~\ref{QCPSchematic} to a new type depicted in
Fig.~\ref{MQCPSchematic}. Here, the quantum critical point
is turned into the starting point (or end point) of the quantum critical line
which represents topological nature of transitions at $T=0$.
We call this starting and marginal point shown as the cross in Fig.~\ref{MQCPSchematic}, 
the marginal quantum critical point (MQCP), which is completely different from
the conventional quantum critical point.
Then electron correlation effects generate an unconventional universality class of phase transitions,
which is expected neither in GLW category nor in simple topological transitions.

\subsection{Introduction for metal-insulator transitions}
Now we summarize understanding of MI transitions for strongly correlated electrons 
achieved in preceding studies in the literature to make clear our motivation of the present study.   
This issue has attracted
long-standing interest and various important aspects have been clarified~\cite{RMP}, 
although complete understanding has not been reached. 
MI transitions may occur with coupling to lattice distortions
or be accompanied by effects of randomness in real materials.
However,
purely electronic origin of MI transitions
has been a central issue in the strong correlation regime,
where we find many examples called Mott transition.
For this purpose, Hubbard model,
which only considers the local Coulomb interaction on a lattice,
is one of the simplest models to describe MI transitions.
Hubbard model is defined by the Hamiltonian as
\begin{equation}
H=-\sum_{i,j,\sigma}t_{ij}c_{i\sigma}^{\dagger}c_{j\sigma} 
+U\sum_{i}n_{i\uparrow}n_{i\downarrow }
-\mu \sum_{i}N_{i},
\label{HM}
\end{equation}
where  $c_{i\sigma}^{\dagger} (c_{i\sigma})$ creates (annihilates)
an electron with spin $\sigma$ at site $i$, respectively. The number operator is 
$n_{i\sigma}$ and $\mu$ represents the chemical potential.
Hopping integral between site $i$ and $j$ is represented by $t_{ij}$ 
and $U$ denotes the onsite Coulomb repulsion. 
 
Approximate and simple theoretical descriptions of the Mott transition 
at $T=0$ without any symmetry breaking being involved was proposed in seminal works
of Hubbard~\cite{Hubbard} and Brinkman and Rice~\cite{Rice}.  
These two works offer completely different pictures for the transition.
Hubbard approximation offers a picture of simple splitting of a band into
an upper and a lower band with a gap 
between these two bands. The transition to Mott insulator occurs simply because the chemical potential
moves into the gap.  The ``quasiparticle" weight $Z$ at the Fermi level is retained nonzero even at
the transition, although the ``large Fermi surface" with the Luttinger volume is not 
preserved and a small pocket of the Fermi surface shrinks to zero at the transition.  
Brinkman-Rice picture obtained from the Gutzwiller approximation, in contrast, 
preserves the Luttinger volume
and the Mott insulator is realized by the vanishing quasiparticle weight $Z$.  
Dynamical mean-field theory(DMFT)~\cite{DMFT} proposed some unification of these two pictures
with the formation of the upper and lower Hubbard bands reproduced together with the vanishing $Z$ in the coherent 
additional ``band".  In DMFT results, the Mott transition itself is triggered by the vanishing coherent band  and the 
criticality is ultimately determined from that of the Brinkman-Rice scenario.

Recently, DMFT has been improved by taking into account the momentum dependence of the self-energy 
in the correlator projection method~\cite{Onoda}.  The Mott transition obtained from this improvement
shows that a transition qualitatively different from the original DMFT scenario occurs, 
where the renormalization factor 
$Z$ is kept nonzero until the transition~\cite{Hanasaki}. Instead of vanishing $Z$, 
the transition is realized by shrinkage and vanishing of the Fermi-surface
pocket with some similarity to the Hubbard picture.  A crucial difference from the Hubbard picture is that 
the upper and lower Hubbard bands may have qualitatively different dispersions accompanied by electron differentiation in momentum space~\cite{Onoda,Differentiation1,Differentiation2}.
The vanishing pocket looks to show very anisotropic spectral weight with strong intensity on 
the inner-half circle and weak intensity on the outer-half circle, making an apparent 
arc-type structure for the model on the square lattice.  This arc-type structure
is consistent with the experimental observations in the copper oxides obtained from angle resolved 
photoemission spectra (ARPES)~\cite{ARPES, Armitage}.  
This Fermi-surface pocket appears after the Lifshitz transition within the metallic phase, 
where the topology
of the Fermi surface changes from that of a large Fermi surface to pockets.  
Such Lifshitz transitions seem to inevitably occur, which invalidates
the Luttinger theorem after the transition to the pockets~\cite{Hanasaki}.  
A similar conclusion was obtained by cellular DMFT study for systems 
with a one-dimensional anisotropy~\cite{Georges} as well as in 2D~\cite{Civelli}, 
and by dynamical cluster approximation for 2D Hubbard model~\cite{MaierJarrell}, 
where the momentum dependence of the self-energy was considered in different ways.
The serious modification of DMFT and the complete change in the nature of the MI transition 
with vanishing pocket of the Fermi surface seems to have a deep connection
to the satisfaction of the hyperscaling and a momentum-space differentiation as we discuss below.

A picture of MI transitions completely different from various mean-field approximations 
including DMFT has been proposed from the 
scaling theory of Mott transition~\cite{Imada_94,Imada_95}, where the singularity of the free energy $F$ at the 
Mott transition is assumed to follow the hyperscaling form 
  \begin{equation}
    F\propto \xi^{-(d+z)}.
    \label{hyperscaling}
  \end{equation}
Here, $\xi$ in the metallic side is the correlation length defined by $X^{-1/d}$ 
for the metallic carrier concentration $X$ and spatial dimensionality $d$, 
while it is the localization length of carriers in the insulating side.
More precise definition of $\xi$ should be given from the correlation length of 
the order-parameter correlation function, which will be discussed 
in the end of this section and will be clarified 
that the above relation is correct.  
The hyperscaling is normally satisfied below the upper critical dimension $d_u$, 
whereas the mean-field theories are satisfied above $d_u$.  
Therefore, the scaling theory and various mean-field theories are incompatible in general.

Before proceeding to more involved discussions on the scaling theory, we briefly summarize 
basic points of critical phenomena and universality classes particularly 
for metal-insulator transitions used in the later discussions. 
The universality class of phase transitions is characterized by critical exponents.
We first discuss relations for critical exponents which is applicable 
irrespective of the validity of either the hyperscaling or mean-field approximations.
When we take the carrier concentration $X$ as the order parameter in the metallic side, 
the critical exponent $\delta$ is defined by 
   \begin{equation}
    X\propto \zeta^{1/\delta},
    \label{delta}
  \end{equation}
which measures the scaling behavior of 
the growth of the order parameter as a function of the field $\zeta$,
which is conjugate to the order parameter $X$. Namely, $\zeta$ is either 
the chemical potential $\mu$ for the filling-control transition or $U/t$ 
in the Hubbard model for the band-width control transition.  In the latter case, $\zeta$ is, for example,  
controlled by pressure $P$ in actual experiments.
The carrier concentration $X$ is the doping concentration itself for the filling-control case while
it is the unbound doublon (doubly occupied sites) and holon (empty site) concentrations 
measured from the values at the critical point for the band-width control case.

The exponent $\nu$ is defined from the relation of the control parameter $g$ to $\xi$ 
as    
   \begin{equation}
    \xi \propto |g-g_c|^{-\nu}.  
     \label{nu}
     \end{equation}
Here, the control parameter $g$ may also be controlled by $U/t$ or $\mu$, which makes some complexity and could cause a confusion,
because these two quantities can control both of $\zeta$ and $g$. 
This is somehow in contrast with the simple case of Eq. (\ref{GLFreeEnergy}), where $h$ and $a$ may be 
controlled independently, say, by the magnetic field and temperature, respectively.   
In the later microscopic description in Sec. \ref{sec:Exponents}, this somewhat confusing situation will be resolved.   

The exponent $\gamma$ is defined from the order-parameter susceptibility $\chi=dX/d\zeta$ from the scaling
   \begin{equation}
     \chi\propto |g-g_c|^{-\gamma}.
    \label{gamma}
  \end{equation}
  The jump of $X$ at the first-order boundary grows from the critical point as
   \begin{equation}
    X \propto |g-g_c|^{\beta},
     \label{beta}
  \end{equation}
The exponent $\alpha$ is obtained from the second derivative of the free energy as 
   \begin{equation}
d^2 F/dg^2\propto |g-g_c|^{-\alpha}.   
\label{alpha}
  \end{equation}
It should be noted that $\alpha$ for quantum phase transitions 
does not express the exponent for the specific heat 
as in thermal transitions.
The set of these exponents has a significant meaning 
because the set specifies the fundamental nature of the phase 
transitions called the universality class.

Conventional scaling laws such as Widom's law
   \begin{equation}
    \beta(\delta-1)=\gamma
    \label{Widom}
  \end{equation}
and Rushbrooke's law
   \begin{equation}
    \alpha+2\beta+\gamma=2
    \label{Rushbrooke}
  \end{equation}
may be satisfied for these exponents.
So far, the control parameter $g$ is not fixed yet and one may have several choices.
Some of the exponents need trivial transformation depending 
on the definition of $g$~\cite{comment}.

When the hyperscaling assumption is employed, 
the length scale which diverges toward the critical point is unique, 
and it must be the mean carrier distance given by    
   \begin{equation}
       \xi\propto X^{1/d}.
    \label{xi}
  \end{equation} 
Then the free energy expressed by Eq. (\ref{hyperscaling})
leads to 
   \begin{equation}
    F\propto X^{(d+z)/d}.
    \label{hyperscaling2}
  \end{equation}
This is the form at $\zeta=0$ and $g-g_c=0$. 
Scaling relations are derived by further adding the term of the ``chemical potential" $\zeta$ 
and the term of coupling to $g-g_c$ as
   \begin{equation}
    F= -\zeta X + B_0(g-g_c)X^{\phi}+ C X^{(d+z)/d},
    \label{hyperscaling3}
  \end{equation}
where $B_0$ and $C$ are constants and 
$\phi$ should be separately given from a physical consideration or microscopic derivation of the transition.
Note that the definition of $g$ is now fixed as the quantity to control the transition at $\zeta=0$. 
The scaling between $X$ and $\zeta$, namely, Eq. (\ref{delta}) is obtained by putting $g-g_c=0$ 
in Eq. (\ref{hyperscaling3}) and minimize the free
energy by $dF/dX=0$. This yields 
   \begin{equation}
    \delta=z/d.
    \label{delta2}
  \end{equation}
The scalings between $X$ and $g$, namely, Eqs. (\ref{gamma}) and (\ref{beta}) are obtained by putting $\zeta=0$ 
in Eq. (\ref{hyperscaling3}) and minimize the free energy by $dF/dX=0$. This yields 
   \begin{equation}
   \beta=\frac{1}{1-\phi+z/d},    
   \label{beta2}
  \end{equation}
and 
     \begin{equation}
    \gamma=\frac{z-d}{z+d(1-\phi)}.     
      \label{gamma2}
      \end{equation}
From (\ref{nu}), (\ref{beta}) and (\ref{xi}), 
a further relation   
   \begin{equation}
     \nu=\beta/d    
    \label{nu2}
    \end{equation} 
is obtained.
Note that the scaling laws Eqs. (\ref{Widom}) and (\ref{Rushbrooke}) are obviously satisfied 
for Eqs. (\ref{delta2})-(\ref{nu2}).
From the scalings (\ref{hyperscaling}), (\ref{nu}) and (\ref{xi}),
we obtain $F\propto |g-g_c|^{\nu (d+z)}$ for nonzero $g-g_c$. From this and Eq. (\ref{alpha}), the Josephson relation    
\begin{equation}
    \alpha=2-\nu(d+z)
    \label{Josephson}
  \end{equation}
is obtained. 
In addition, Drude weight is shown to follow~\cite{RMP} 
   \begin{equation}
    D\propto X^{1+(z-2)/d}.
    \label{Drude}
   \end{equation}

To estimate accurate exponents of microscopic models, in principle, one needs to go beyond the mean-field approximation, 
because the reliability of the mean-field approximation itself has to be critically tested. 
Available theoretical tools for this purpose is severely limited, because of various difficulties
in theoretical approaches for strongly correlated electrons. In the literature, there exist only small number of available
estimates from unbiased approaches such as the quantum Monte Carlo and path-integral renormalization group
methods.  For the filling-control transition of Hubbard model or the $t$-$J$ model on a square lattice, 
by using the quantum Monte Carlo method~\cite{Furukawa_0,Furukawa_1,Furukawa_2}
and power Lanczos method~\cite{Kohno}, the charge compressibility has been shown to follow 
the scaling $dX/d\mu \propto X^{-1}$, from which $\delta$ is estimated from Eq. (\ref{delta}) as 
$\delta\sim 2$. The result obtained from the path-integral renormalization group method is 
also consistent with this scaling~\cite{Watanabe}.  When the hyperscaling holds, $\delta=2$ and Eq. (\ref{delta2})
in two dimensions leads to $z=4$.  On the other hand, simple GLW scheme always gives $\delta>3$ and is incompatible with
the numerical results.  In fact, the exponents of the Ising model are $\delta\sim4.8$ for three-dimensional systems  
and $\delta= 15$ in two dimensions. Any other types of symmetry breakings indicate $\delta>3$.
Note also that the degeneracy temperature (effective Fermi temperature) $T_F$ is scaled by $T_F\propto \xi^{-z} 
\propto X^{z/d}$ resulting in $T_F\propto X^2$ for $z=4$ and $d=2$, if the hyperscaling is satisfied~\cite{Imada_95}.

Experimentally, the exponent $\delta$ has been examined for filling-control transitions 
by measuring the chemical potential shift with increasing doping
by photoemission studies in several high-$T_c$ copper oxides~\cite{Fujimori,Fujimori2}.  In particular,
Bi$_2$Sr$_2$CaCu$_2$O$_{8+y}$ and (La,Sr)$_2$CuO$_4$ show pinnings of the chemical potential at low 
doping, indicating an enhancement of the charge compressibility and consistency with $\delta=2$.
$^3$He adsorbed on a graphite surface, offering a unique two-dimensional fermion system with strong correlation 
effects, appears to show an anomalous suppression of $T_F$ near the commensurate solid phase in accordance with the trend of
$T_F\propto X^2$ consistently with $z=4$~\cite{Saunders} with the hyperscaling being satisfied.

In addition, the Drude weight has been calculated to be consistent with
$D\propto X^2$~\cite{Tsunetsugu,Nakano} which suggests $z=4$ from the above relation (\ref{Drude})
if the hyperscaling is satisfied.  These two independent estimates of $z$ coincide each other and support
the hyperscaling.

Other exponents are directly related to $g$ dependence of $F$ and it is fixed when $\phi$ is specified.
If one put $z=4$ and $d=2$,
\begin{eqnarray}
\nu &=& \frac{1}{2(3-\phi)}, \label{nu3}\\
\alpha &=& 2-3\frac{1}{3-\phi}, \label{alpha3}\\
\beta &=& \frac{1}{3-\phi}    
\label{beta3}
  \end{eqnarray}
and 
\begin{equation}
\gamma=\frac{1}{3-\phi}    
\label{gamma3}
  \end{equation}
are obtained.  

This unconventional universality class has further been analyzed and $\phi=2$ is obtained from a plausible assumption 
on the quasiparticle dispersion~\cite{MQCP_3,MQCP_2,MQCP}.
Then, 
the universality class is proposed to be characterized by a set of the exponents 
  \begin{equation}
     z=4, \alpha=-1, \beta=1, \gamma=1, \delta=2, \nu=1/2 \ \ {\rm and} \ \ \eta=0.    
    \label{Exponents}
  \end{equation}

At finite temperatures, it has been theoretically proposed
that the MI transition is equivalent to the gas-liquid transition and
its universality class belongs
to the Ising universality class~\cite{DMFT_Landau, Castellani}.
Experimentally, from recent careful study of V$_{2}$O$_{3}$ by Limelette $et$ $al$~\cite{V2O3_S},
they have obtained the critical exponents
which are consistent with that of the 3D Ising model,
namely, $\beta\sim 0.34$, $\delta\sim 5.0$ and $\gamma\sim 1.0$ in the region very close to the critical point
whereas they have obtained consistency with
the mean field exponents for the Ising model in the most part away from the critical point. 
On the contrary, 
recent experimental results on $\kappa$-(ET)$_{2}$Cu[N(CN)$_{2}$]Cl~\cite{Kanoda_MI,Kanoda_MI_Nature}
indicate the critical exponents $\beta=1, \gamma=1$ and $\delta=2$, which are consistent with Eq. (\ref{Exponents})
and indicates an evidence for the relevance of the present universality.


	Before closing this section, we remark on the definition
	of $\xi$ and the dynamical exponent $z$. 
	In the case of MI transitions, correlation functions
	do not decay exponentially as a function of distance in the metallic phase. 
	Therefore, the conventional definition of $\xi$ given in Eq. (\ref{eq:con_xi})
	is not applied.  
	For example, noninteracting electrons have a density-density correlation function 
	$\langle X(r)X(0) \rangle$
	which asymptotically decays as $1/r^3$ with Friedel oscillations in two-dimensions~\cite{Furukawa_1}. 
	Such an asymptotic decay crossovers to a constant at shorter distance, 
	where the crossover length is determined from the inverse of Fermi wave number $k_F$. 
       Since $1/r^3$ decay is absolutely convergent in the integration, the integration over distance is contributed
       dominantly from the region inside of the
       length scale of $k_F^{-1}$. 
	From this, $\xi$ is given by $\xi\propto k_F^{-1}\propto X^{-1/d}$.
	More precisely, the correlation length $\xi$ is defined by 
	\begin{align}
	\xi \equiv \sqrt[\displaystyle d]{\frac{\int {\rm d}r \langle X(r,0)X(0,0) \rangle}
	{\langle X(0,0)X(0,0) \rangle}}. 
	\end{align}
	In the similar way, the temporal correlation length $\xi_{t}$
	is defined by 
	\begin{align}
	\xi_{t} &=\frac{\int {\rm d}t \langle X(0,t)X(0,0) \rangle}{\langle X(0,0)X(0,0) \rangle}. 
	\end{align}
	Then the dynamical exponent $z$ is obtained from
	\begin{align}
		z=\lim_{g\rightarrow g_{c}} \frac{\log{\frac{\chi}{S}}}{\log{\xi}},
	\end{align}
	where we have used relations for the equal-time structure factor $S$ and susceptibility $\chi$,
	given by
	\begin{eqnarray}
	S & \equiv & \int dr \langle X(r,0)X(0,0) \rangle \propto \xi^{-d}\label{eqn:Schi2}\\
	\chi & \equiv & \int drdt \langle X(r,t)X(0,0) \rangle 
	\propto \xi^{z-d}   \label{eqn:Schi3}
	\end{eqnarray}
	leading to
	\begin{eqnarray}
	\frac{\chi}{S}&\propto& \xi^z. \label{eqn:Schi4}
	\end{eqnarray}

\subsection{Outline of the present work}
	
In this paper, we show that the basic structure of the phase diagram 
for the MI transition involving the quantum critical line and the boundary 
for the first-order transition is correctly captured by the Hartree-Fock theory of the Hubbard model. 
A remarkable result in this paper is that the Hartree-Fock approximation is able to derive 
the above unconventional universality class at MQCP from a microscopic derivation of the
free energy. The obtained universality class turns out to be consistent with the scaling theory.
The mechanism of the emergence of 
MQCP and resultant unconventional universality is clarified from the present microscopic study.
It is ascribed to the selfconsistent positive feedback between metallic carrier density 
and the amplitude of an insulating 
gap already developed in metals. More concretely, increase 
of the preformed gap amplitude decreases the carrier density to lower the kinetic energy, which in turn further enhances the 
preformed gap. On the contrary, the decreasing preformed gap allows more carriers, which further 
reduce the gap to gain the kinetic energy of carriers.  These nonlinear effects cause the first-order transition and MQCP emerges at the border between the 
first-order transition and the quantum critical line.  
Furthermore, the Hartree-Fock theory turns out to remain essentially correct even when one goes beyond
the mean-field approximation and takes into account quantum fluctuations.  This is because the 
transition always occurs as that at the upper critical dimension irrespective of the spatial dimensionality. 
We further show that the
Hartree-Fock solution has a finite-temperature crossover to the conventional
Ising universality class.  
We discuss implications and consequences of this unexplored type of universality class emerging near MQCP
sandwiched by GLW critical and quantum critical lines.
We argue that low-energy excitations are described neither by the original single-particle picture
nor by the bosonic density fluctuations when the both characters of transitions meet.

We also discuss comparison of the present result with the experimental indications obtained so far.
It is shown that the critical exponents at MQCP together with the finite-temperature crossover 
are completely consistent with the whole aspects of the experimental results.

A part of the discussions in this paper is already briefly given~\cite{MisawaYamajiImada}.
In this paper,
we present the results of quantum MI transitions in greater detail.
In particular, we show details of the procedure to obtain
the non-GLW free-energy expansion and critical exponents together with thorough discussions on implications.

The organization of this paper is as follows:
In Sec. \ref{sec:HFA}, we show details of our Hartree-Fock approximation.
In Sec. \ref{sec:Exponents}, we present Hartree-Fock solutions of the MI transition by introducing the 
preexisting order parameter.  It captures some essence of the unconventional universality
class of the transition from metals to insulators caused by the electron correlation effects.         
Using the free-energy expansion, we analytically obtain the critical exponents of quantum MI transitions,
which belongs to an unconventional universality. 
We also discuss implications beyond the mean-field approximations in Sec. \ref{sec:Exponents}.
In Sec. \ref{sec:NumericalT0}, we show numerical results of Hartree-Fock approximation at $T=0$, which are completely
consistent with the analytical results in Sec.\ref{sec:Exponents}.
In Sec. \ref{sec:NumericalT1}, finite-temperature crossover is calculated.  It shows a well-defined crossover from 
the Ising universality in the close vicinity of the critical point to the quantum universality governed by MQCP when
the parameter deviates from the critical point.
Section \ref{sec:Comparison} describes comparisons of the present implications with relevant experimental results.
Section \ref{sec:Summary} is devoted to summary and discussions.  

\section{Hartree-Fock approximation}
\label{sec:HFA}
In this paper, to capture the essence of quantum MI transitions,
we extend Hubbard model (\ref{HM}) by considering the nearest-neighbor repulsion term proportional
to $V$ on the $N_{s}=L\times L$
square lattice.  It is defined by 
\begin{align}
H&=\sum_{k,\sigma}(\xi_{1}({\bf k})+\xi_{2}({\bf k}))c_{k\sigma}^{\dagger}c_{k\sigma} \notag \\
&+U\sum_{i}n_{i\uparrow}n_{i\downarrow }+V\sum_{\langle ij\rangle}N_{i}N_{j}
-\mu \sum_{i}N_{i},
\label{EHM}
\end{align}
where $N_{i}=(n_{i\uparrow}+n_{i\downarrow})$ are number operators.
We restrict the hopping terms to the nearest neighbor pairs given by
the dispersion 
$\xi_{1}({\bf k})=-2t(\cos{k_{x}}+\cos{k_{y}})$ and that of the
next-nearest neighbor pairs by
$\xi_{2}({\bf k})=4t^{\prime}\cos{k_{x}}\cos{k_{y}}$, respectively.

Using this extended Hubbard model,
we study quantum MI transitions in two-dimensional system
within the Hartree-Fock approximation.
Although the Hartree-Fock analyses have been done for the Hubbard model
in the literature~\cite{KondoMoriya, Vollhardt}, the criticality of MI transitions
has not been examined. By considering the MI transitions in the
ordered phase, we are able to capture a remarkable aspect of MI transitions
from a microscopic model.

To consider the MI transitions within the Hartree-Fock approximation,
we consider a symmetry broken long-range order
such as charge or antiferromagnetic (AF) ordering
with the periodicity commensurate to the lattice.
Then we study the MI transitions in the symmetry 
broken phase.
We note that essence of the MI transitions does not depend on
the type of the ordering.
For the extended Hubbard model defined in Eq. (\ref{EHM}),
two simple orders become stable.
When $4V>U$,
the extended Hubbard model 
has a tendency to charge ordering, where the empty
and doubly occupied sites alternatingly align within the 
mean-field level. 
The order parameter $m$ of checker-board-type doublon alignment is
defined by  
\begin{equation}
m=\frac{1}{N_{s}}\sum_{i}\langle (n_{i\uparrow}+n_{i\downarrow})\exp(i\mathbf{Q}\mathbf{r_{i}})\rangle,
\end{equation}
where ${\bf Q}$ is the ordering wave vector $(\pi,\pi)$.
For $U>4V$, 
the antiferromagnetic ordering becomes stable.
The order parameter of the antiferromagnetic ordering
$m_{\rm AF}$ is defined by
\begin{equation}
m_{\rm AF}=\frac{1}{N_{s}}
\sum_{i}\langle (n_{i\uparrow}-n_{i\downarrow})\exp(i\mathbf{Q}\mathbf{r_{i}})\rangle.
\label{eq:m_AF}
\end{equation}

Hereafter, we mainly consider the charge ordering. However, within the mean-field approximation, 
the antiferromagnetic order and charge order play the same role on MI transition and 
by replacing the order parameter $m$ with the antiferromagnetic one, one can easily reach the same results 
for the nature of the MI transition.

Here we take the mean field as
\begin{equation}
\langle n_{i\uparrow}\rangle=\langle n_{i\downarrow}\rangle=\frac{(n+m\exp(i\mathbf{Q}\mathbf{r_{i}}))}{2},
\end{equation} 
with $n$ being the average  
charge density given by $1/N_{s}\sum_{i}\langle n_{i\uparrow}+n_{i\downarrow}\rangle$.
Using this mean field, we decouple the interaction term.

The on-site interaction term is decoupled as 
\begin{eqnarray}
U\sum_{i}n_{i\uparrow}n_{i\downarrow}  
&\sim& U\sum_{i}\left(n_{i\uparrow}\langle n_{i\downarrow}\rangle+n_{i\downarrow}\langle n_{i\uparrow}\rangle
-\langle n_{i\uparrow}\rangle\langle n_{i\downarrow}\rangle\right) \nonumber \\ 
&=&\frac{U}{2}\sum_{i\sigma}\left(n_{i\sigma}\left(n+me^{i\mathbf{Qr_{i}}}\right)\right) \nonumber \\
&-&\frac{UN_{s}\left(n^2+m^2\right)}{4}. 
\label{Unn}
\end{eqnarray}
The nearest-neighbor interaction term is decoupled as 
\begin{eqnarray}	
	V\sum_{\langle ij\rangle}N_{i}N_{j} 
    &\sim& V\sum_{\langle ij\rangle}(N_{i}\langle N_{j}\rangle
	+N_{j}\langle N_{i}\rangle-\langle N_{i}\rangle\langle N_{j}\rangle) \nonumber \\
  &=&4nV\sum_{i}N_{i}-4mV\sum_{ i}N_{i}\exp(i\mathbf{Qr_{i}}) \nonumber \\
  &-&2VN_{s}(n^2-m^2). 
\label{VNN}
\end{eqnarray}

Finally we obtain a Hartree-Fock Hamiltonian($H_{HF}$) in the momentum space as 
\begin{eqnarray}
 H_{HF}&=&\sum_{k,\sigma}\left(\epsilon({\bf k})
 +\frac{Un}{2}+4nV-\mu \right)c_{\mathbf{k} \sigma}^{\dagger}c_{\mathbf{k} \sigma} \nonumber \\
	&+&\left(\frac{U}{2}-4V\right)m\sum_{\mathbf{k} \sigma}c_{\mathbf{k+Q} \sigma}^{\dagger}c_{\mathbf{k} \sigma} \nonumber \\	
 	&-&\left(\frac{UN_{s}(n^2+m^2)}{4}+2VN_{s}(n^2-m^2)\right),
\label{HHF}
\end{eqnarray}
where we define $\epsilon({\bf k})=\xi_{1}({\bf k})+\xi_{2}({\bf k})$.

Diagonalizing the Hamiltonian leads to two bands of the form
\begin{align}
&E_{\pm}=\xi_{2}({\bf k})\pm\sqrt{\xi_{1}({\bf k})^{2}+\Delta^2}-\mu, 
\label{eq:Band}
\end{align}
where $\Delta=mg$ and we now take $g=4V-\frac{U}{2}$ as the control parameter,
because in this section we mainly consider the band-width control transition.
In Eq.~(\ref{eq:Band}), we dropped a constant term $n(g+U)$.

Using this Hartree-Fock band dispersion we obtain the free energy 
\begin{align}
 F_{HF}&=-\frac{T}{N_{s}}\log Z_{HF} \notag \\  
	&=-\frac{2T}{N_{s}}\sum_{\mathbf{k}, \eta}
\log\left(1+\exp\left(-\frac{E_{\eta}\left(\mathbf{k}\right)}{T}\right)\right) \notag \\ 
	&-\left(\frac{U(n^2+m^2)}{4}+2V(n^2-m^2)\right), \label{eq:HF_F}
\end{align}
where the suffix $\eta$ takes $\pm$.

From this free energy, the order parameter $m$ is determined from the
self-consistency condition 
\begin{align}
	\frac{1}{g}=&\frac{2}{N_{s}}\sum_{\mathbf{k}} 
	\frac{f(E_{-}(\mathbf{k}))-f(E_{+}(\mathbf{k}))}{E_{+}(\mathbf{k})-E_{-}(\mathbf{k})}, 
        \label{eq:HF_self}
\end{align}
where $f(x)$ is the Fermi-Dirac distribution function.
The particle density $n$ is given by
\begin{equation}
	n=\frac{1}{N_{s}}\sum_{\mathbf{k}} f(E_{-}(\mathbf{k}))+f(E_{+}(\mathbf{k})). 
     \label{eq:HF_density}
\end{equation}
Details of numerical calculations of Eqs. (\ref{eq:HF_self}) and (\ref{eq:HF_density})
are shown in Appendix \ref{ap:A}.  
	
It turns out that the self-consistent equation (\ref{eq:HF_self}) with 
Eq. (\ref{eq:Band}) for the charge ordering  is equivalent
to the self-consistent equation for the AF ordering defined 
in Eq. (\ref{eq:m_AF})~\cite{KondoMoriya, Vollhardt}.
The equivalence holds by replacing the charge order parameter
$m$ with the antiferromagnetic(AF) order $m_{\rm AF}$
and by putting  $g=U/2$. 
However, one should be careful about this mapping.
This mapping is complete only on the Hartree-Fock level.
The charge ordering at the $(\pi, \pi)$ periodicity
is the consequence of the discrete symmetry breaking 
while the AF ordering is realized by the continuous symmetry breaking of $SU(2)$ 
symmetry. Therefore, if it would be exactly solved  
in two-dimensional systems, the charge ordering can indeed exist at finite temperatures, though 
AF ordering cannot exist at finite temperatures by Mermin-Wargner theorem.
Therefore, the Hartree-Fock approximation captures 
some essence of the  charge ordering at finite temperatures,
while the AF ordering at nonzero temperatures in two dimensions contains 
an artifact of the Hartree-Fock approximation, although the development of the charge gap itself
is not an artifact, which leads to the Mott gap formation even without any long-range order.
The physics of unconventional quantum criticality, which we clarify later, 
emerges from the gap formation itself and therefore the essence is captured 
in the Hartree-Fock calculation.

\section{Free-energy expansion and critical exponents}
\label{sec:Exponents}
\begin{figure}[h!]
	\begin{center}
		\includegraphics[width=7cm,clip]{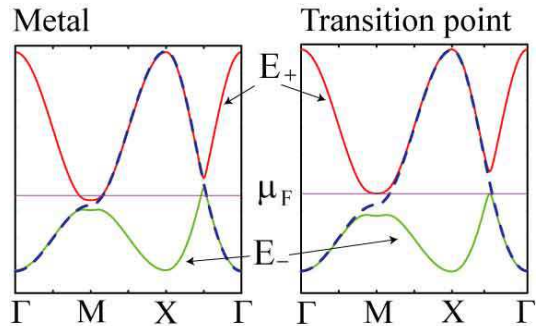}   
	\end{center}
	\caption{(color online) Quasiparticle band structures of the Hartree-Fock solution 
       for metallic phase (left) and the continuous MI transition
	point (right) are shown. Broken (blue) line shows the free-particle band dispersion defined
	by $\epsilon({\bf k})=-2t(\cos{k_{x}}+\cos{k_{y}})+4t^{\prime}\cos{k_{x}}\cos{k_{y}}$.
	The definition of the upper (lower)  band is defined in Eq. (\ref{eq:Band}).
	Fermi energy $\mu_{F}$ is represented by the horizontal line. In the metallic phase,
       the bottom of the upper band at M point ($(\pi,0)$ and its equivalent points) is lower than 
       the top of the lower band at $(\pi/2,\pi/2)$ and its equivalent points 
       so that the indirect insulating gap is closed.}
	\label{fig:Band}
	\end{figure}

        In this section, we analyze singularities of the free energy around
        the MI transition within the Hartree-Fock approximation
        at $T=0$ and obtain the critical exponents of the transition.
        When the order parameter $m$ and the gap $\Delta = mg$ are sufficiently small near 
        the charge-order or antiferromagnetic transition, the insulating gap does not open
        for nonzero positive $t^{\prime}$. 
        This is because, the upper band $E_+$ and lower band $E_-$ overlap for a small gap. 
        From Eq. (\ref{eq:Band}), the top[bottom] of the lower[upper] band is
	 located at $(k_{x}, k_{y})=(\pi/2, \pi/2)\left[(\pi, 0)\right]$, respectively.
	 The continuous MI transition occurs when the top of
	 the lower band reaches the same energy as that of the bottom of the upper band
	 as shown in Fig.~\ref{fig:Band}.
	 Therefore, the condition of the continuous MI transition is given by
	 \begin{equation}
		E_{+}(\pi, 0, \Delta_{c})=E_{-}(\pi/2, \pi/2, \Delta_{c}),
	 \end{equation}
	 where $\Delta_{c}$ is the critical gap of the continuous MI transition.
	 From this, at $T=0$, we obtain the critical gap $\Delta_{c}$ as
	 \begin{equation}
        \Delta_{c}=2t^{\prime}.
	 \end{equation}

        This quantum MI transition
        occurs between charge-ordered metal and charge-ordered insulator,
        thus the symmetry does not change at the transition.
        It is legitimate to expand the free energy around
        nonzero order parameter measured from the critical point of the MI transition.
        One might speculate that the free energy expressed only by the order parameter $\Delta$
        in Eq. (\ref{eq:HF_F}) would be regular around the MI transition.
        However, as we will see below, this expansion has a piecewise analytic form in the metallic and insulating
        phases separately with a jump of the expansion coefficient at the transition point. 
	 This somewhat unconventional form of expansion is derived from the fact that some of the expansion 
        coefficients explicitly depend on the density of states at the Fermi level, which has a discontinuity
        at the MI transition.  
        This free-energy structure with a jump is in a complete contrast with the GLW form.  

	 In this section, we consider the band-width control transition at half filling ($n=1$)
	 in the canonical ensemble.  
        In the scheme of the canonical ensemble, we add a term $\mu n$ to the free energy 
	 defined in Eq. (\ref{eq:HF_F}). 
	 The Hartree-Fock free energy $F_{CHF}$ is given by
	 \begin{equation}
		F_{CHF}=-\frac{2T}{N_{s}}\sum_{\mathbf{k}, \eta}
             \log\left(1+\exp\left(-\frac{E_{\eta}\left(\mathbf{k}\right)}{T}\right)\right)
             +\frac{\Delta^2}{2g}+\mu n,
         \label{eq:CHF}
	 \end{equation}
        where $\eta$ runs over $+$ and $-$.
        Hereafter, we describe $F_{CHF}$ as $F$.
	 

	 \subsection{Free-energy expansion}
	 
	 Since $\Delta$ dependence of $F$ is explicitly given in 
        Eq. (\ref{eq:CHF}), it allows a formal expansion of $F$ in terms of $\Delta$ around 
        a nonzero value $\Delta_c$, by taking an expansion parameter $Y=\Delta_c-\Delta$.  
        The region $Y>0$ ($Y<0$) represents the metallic
	 phase (insulating phase), respectively.
	 The form of expansion is given as 
	\begin{align}
		F({\Delta})&=F(\Delta_{c})+AY+
		\frac{B}{2!}Y^{2}+\frac{C}{3!}{\Delta}Y^{3}
		+\dots \label{eq:MQCP_GL}, \\ 
	\end{align}
	where the coefficients are given by
   	\begin{align}
		A&=-\frac{\partial F}{\partial \Delta}\Bigr|_{\Delta=\Delta_{c}}
		=-\left(\frac{1}{g}-W_{1} \right)\Delta_{c},  \label{eq:A}\\
		B&=\frac{\partial^{2} F}{\partial \Delta^{2}}\Bigr|_{\Delta=\Delta_{c}} \notag \\
		&=\frac{1}{N_{s}}\sum_{\bf{k}, \eta}\Biggl(
		\frac{\partial f(E_{\eta}({\bf k}, \Delta))}{\partial \Delta}
		\frac{\Delta}{ \eta\sqrt{\xi_{1}^{2}+\Delta^{2}}}\Biggr)\Biggr|_{\Delta_{c}}  
		-W_{2}, \label{eq:B} \\  
		C&=-\frac{\partial^{3} F}{\partial \Delta^{3}}\Bigr|_{\Delta=\Delta_{c}} \notag \\
		&=-\frac{1}{N_{s}}\sum_{\bf{k}, \eta}\Biggl(
		\frac{\partial^{2} f(E_{\eta}({\bf k}, \Delta))}{\partial \Delta^{2}}
		\frac{\Delta}{ \eta\sqrt{\xi_{1}^{2}+\Delta^{2}}}\Biggr)\Biggr|_{\Delta_{c}}-3W_{3},
		 \label{eq:C}
	\end{align}
    with the definitions  
    \begin{align}
		W_{1}&=v_{1}, \\
		W_{2}&=v_{1}-\Delta_{c}^2v_{3}, \\
		W_{3}&=v_{3}\Delta_{c}-v_{5}\Delta_{c}^{3}. 
	\end{align}
	Here, $v_{n}$ is defined by
	\begin{equation}
		v_{n}=-\frac{1}{N_{s}}\sum_{{\bf k}, \eta}\eta(\xi_{1}^2+\Delta_{c}^{2})^{-\frac{n}{2}}
		f(E_{\eta}({\bf k}, \Delta_{c})).
	\end{equation}
    To derive  Eqs. (\ref{eq:B}) and (\ref{eq:C}), we neglect terms those are proportional to
	\begin{equation}
	 \frac{\partial f(E_{\eta}({\bf k}, \Delta))}{\partial \Delta}
	 \frac{\partial }{\partial \Delta}(\frac{\Delta}{\sqrt{\xi_{1}^{2}+\Delta^{2}}})\Biggr|_{\Delta_{c}}, \label{eq:neg} 
	\end{equation}
       because of the following reason:
	At $T=0$, this term  has nonzero value only at $E_{\eta}({\bf k}, \Delta)=0$ 
	(on the Fermi surface) because $\partial f(E_{\eta}({\bf k} ,\Delta))/\partial \Delta$ 
	becomes the delta function.
	At the MI transition point, since the Fermi energy
	has the same value as that of the top [bottom] of the lower [upper] band,
	$\partial f(E_{-}({\bf k},\Delta))/\partial \Delta$
	[$\partial f(E_{+}({\bf k},\Delta))/\partial \Delta$]
	 has nonzero value only at $(\pi/2,\pi/2)$[$(\pi,0)$], respectively.
	However, at the MI transition point,
	namely at $(\pi/2,\pi/2)$ and $(\pi,0)$, $\xi_{1}=0$ is satisfied,	
	then $\partial/{\partial \Delta}\left(\Delta/\sqrt{\xi_{1}^{2}+\Delta^{2}}\right)$ becomes zero.  	  
	From this, the term defined in Eq. (\ref{eq:neg}) vanishes
	at the MI transition point $\Delta=\Delta_{c}$ at $T=0$.	

	Next, we derive more explicit forms of coefficients $B$ and $C$ at $T=0$. 
       We first consider the contribution of the first term of Eq. (\ref{eq:B}) and (\ref{eq:C})
	at $T=0$.
	The first term of Eq. (\ref{eq:B}) is given by
	\begin{align}
		&\frac{1}{N_{s}}\sum_{\bf{k},\eta}\Biggl(
		\frac{\partial f(E_{\eta}({\bf k}, \Delta))}{\partial \Delta}
		\frac{\Delta}{ \eta\sqrt{\xi_{1}^{2}+\Delta^{2}}}\Biggr)\Biggr|_{\Delta_{c}} \notag \\
		&=-\sum_{\eta}\int_{-\Lambda}^{\Lambda}{\rm d}E D_{\eta}(E_{\eta})G(T,E_{\eta}) \notag \\
		&\times\left(\frac{\Delta_{c}}{\eta\sqrt{\xi_{1}^{2}+\Delta_{c}^{2}}}
		+\tilde{\alpha} \right)\left(\frac{\Delta_{c}}{\eta\sqrt{\xi_{1}^{2}+\Delta_{c}^{2}}} \right),
		 \label{eq:B_1}
 	\end{align}
	where $D_{\eta}(E_{\eta})$ is the density of states of the upper (lower) band
    for $\eta=+1 (-1)$, respectively, $\Lambda$ is a cut-off 
	and  $G(T,E)$ is defined by
	\begin{align}
	    G(T,E)&=\frac{\exp(E/T)}{T(1+\exp(E/T))^{2}}.
		\label{eq:G}
	\end{align}
	We introduce a constant $\tilde{\alpha}$ for the linear 
	coefficient between $\Delta_{c}-\Delta$ and $\mu$ as
	\begin{align}
		\mu&=\tilde{\alpha}(\Delta_{c}-\Delta)-2t^{\prime}. 
		\label{eq:chemi}
   	\end{align}
	Detailed derivation of $\tilde{\alpha}$ at $T=0$ is described in Appendix \ref{ap:B}. 
	Explicit form of $\tilde{\alpha}$ at $T=0$ is given by
	\begin{equation}
		\tilde{\alpha}=\frac{2t^{\prime}-\sqrt{t^{2}-{t^{\prime}}^{2}}}{2t^{\prime}
		+\sqrt{t^{2}-{t^{\prime}}^{2}}}.\label{eq:alpha}
	\end{equation}
	The function $G(T,E)$ becomes the delta function at $T=0$.
	Then, Eq. (\ref{eq:B_1}) is discontinuous at the MI transition at $T=0$,
       because the density of states at the Fermi level is discontinuous between the metallic and 
       insulating sides.  
	In fact, at $T=0$, we obtain the relation
	\begin{align}
		(\ref{eq:B_1}):
		\begin{cases}
		&=-\left(D_{+}(1+\tilde{\alpha})+D_{-}(1-\tilde{\alpha}) \right) 
		 \ \ \text{for $Y>0$} \\
		&=0 \ \ \text{for $Y<0$} 
		\end{cases}	\notag
	\end{align}
	where $D_{+}$($D_{-}$) is the density of states at the bottom (top)
    of the upper (lower) band, respectively. Explicit forms of $D_{+}$ and $D_{-}$ are given by
	\begin{align}
		D_{+}&=\frac{1}{4\pi t^{\prime}} \label{eq:D_u} \\
		D_{-}&=\frac{1}{2\pi\sqrt{t^{2}-{t^{\prime}}^{2}}}.\label{eq:D_l}
	\end{align}
	Detailed derivation of $D_{+}$ and $D_{-}$ are given in
    Appendix \ref{ap:B}. 
	
	Using the similar derivation, we obtain
	the representation of the first term of Eq. (\ref{eq:C}) as
	\begin{align}
		&-\sum_{\eta}\int_{-\Lambda}^{\Lambda}{\rm d}E D_{\eta}(E_{\eta})
		\frac{\partial G(T,E_{\eta})}{\partial \Delta} \notag \\
		&\times\left(\frac{\Delta_{c}}{\eta\sqrt{\xi_{1}^{2}+\Delta_{c}^{2}}}
		+\tilde{\alpha} \right)^2\left(\frac{\Delta_{c}}{\eta\sqrt{\xi_{1}^{2}+\Delta_{c}^{2}}} \right).
		\label{eq:C_1}
	\end{align}
	From this, we obtain the relation as
    \begin{equation}
		(\ref{eq:C_1}):
		\begin{cases}
		&=-(1+\tilde{\alpha})^{2}R_{+}(T)+(1-\tilde{\alpha})^{2}R_{-}(T) \notag \\
		   &+\frac{Q_{+}(T)}{\Delta_{c}}(1+\tilde{\alpha})(3+\tilde{\alpha}) \notag \\
		   &+\frac{Q_{-}(T)}{\Delta_{c}}(1-\tilde{\alpha})(3-\tilde{\alpha})  
		 \ \
		 \text{for $Y>0$} \\
		&=0 
		\ \ \text{for $Y<0$} 
		\end{cases}
	\end{equation}
	where $R_{\pm}(T)$ and $Q_{\pm}(T)$ are defined by
	\begin{align}
		R_{\eta}(T)
		&=\int_{-\Lambda}^{\Lambda}dE_{\eta}\frac{\partial D_{\eta}(E_{\eta})}{\partial E_{\eta}}
		G(T,E_{\eta}) \\
	    Q_{\eta}(T)
		&=\int_{-\Lambda}^{\Lambda}dE_{\eta}\frac{D_{\eta}(E_{\eta})^{2}}{D_{\eta}(\sqrt{\xi_{1}^{2}+\Delta^{2}})}
		G(T,E_{\eta}).
	\end{align}

	Finally, at $T=0$,
	we obtain the free-energy expansion for metallic and insulating sides as
	\begin{align}
		F&=AY+\frac{B_{m}}{2!}Y^{2}+\frac{C_{m}}{3!}Y^{3}\dots, 
            \label{eq:ex_free_m}  \\
		F&=AY+\frac{B_{i}}{2!}Y^{2}+\frac{C_{i}}{3!}Y^{3}\dots,
	\label{eq:ex_free_i}
	\end{align}
	where 
	\begin{equation}
	A=-(\frac{1}{g}-W_{1})\Delta_{c}
	\end{equation}
	and the coefficients of the metallic side ($\Delta<\Delta_{c}$) are given as
	\begin{align}
		B_{m}&=-W_{2}-((1+\tilde{\alpha})D_{u}+(1-\tilde{\alpha})D_{l})+\frac{1}{g}\label{eq:B_m} \\
		C_{m}&=-3W_{3}-R_{+}(0)\left(1+\tilde{\alpha}\right)^{2}+R_{-}(0)
		\left(1-\tilde{\alpha}\right)^{2} \notag \\
		&+\frac{Q_{+}(0)}{\Delta_{c}}(1+\tilde{\alpha})(3+\tilde{\alpha})
		+\frac{Q_{-}(0)}{\Delta_{c}}(1-\tilde{\alpha})(3-\tilde{\alpha}),
	\end{align}
	and the coefficients of the insulating side ($\Delta>\Delta_{c}$)  are given as
	\begin{align}
		B_{i}&=-W_{2}+\frac{1}{g} \\
		C_{i}&=-3W_{3}. 
	\end{align}
	We emphasize that, while $A$ is common, the coefficients $B$ and $C$ in the region $Y>0$ have values different from 
       that in the region $Y<0$ at $T=0$.
       Although the coefficients $B$ and $C$ jump at $Y=0$, 
       $F$ is regular (namely, piecewise analytic) within each region $Y>0$ and $Y<0$.
       Then the Taylor series expansion (\ref{eq:ex_free_m}) as well as (\ref{eq:ex_free_i}) is allowed
       in each sector $Y>0$ and $Y<0$ separately and they are unique.  
       The coefficient $C_i$ is negative if $t^{\prime}$ is not too large, 
       because the sign of $W_3$ is determined from the dominant contribution 
       proportional to $v_3$, which has the positive sign due to the contribution from the lower band 
       dominating over the upper-band contribution. 
       On the other hand, $C_m$ is positive for the relevant region of $t^{\prime}$, namely for not too large
       value of $|t^{\prime}|$, 
       because the dominant contribution is then from the term proportional to $Q_+$.  
       Therefore, the MI transition takes place in the parameter regions of $C_m>0$ and $C_i<0$, 
       which makes $F$ always bounded from below within the expansion up to the cubic order of $Y$.
       In fact, we will show below that the expansions up to the cubic order is enough for the 
       purpose of deriving critical properties. 
       Note that this expansion up to the cubic order is consistent with that obtained in the
       scaling theory and its phenomenological extension 
       for two-dimensional systems~\cite{MQCP_3,MQCP_2,MQCP}.
       We will show below that the universality class obtained 
       in the present paper by the microscopic Hartree-Fock calculation indeed agrees 
       with the prediction of the scaling theory and further reveals a microscopic mechanism of the
       emergence of the unconventional universality.

	The carrier density $X$, which is the sum of the electron
	density $X_{-}$ and the hole density $X_{+}$
	in the metallic phase near MI transitions, is given as
	\begin{equation}
		X=X_{+}+X_{-}=\frac{1}{\pi\Delta_{c}}(1+\tilde{\alpha})Y.
          \label{X-Y}
	\end{equation}
	Details of the derivation of Eq. (\ref{X-Y}) are given in Appendix A.
	Since $X$ is proportional to $Y$ in the metallic phase, the singularity
	of $X$ is the same as that of $Y$.
       Therefore, whichever of $X$ or $Y$ is taken as the order parameter, 
       the critical exponents are the same in the metallic side.

	\subsection{Critical exponents}
	
	Using the free-energy expansion obtained in the previous subsection, we
	are able to obtain the critical exponents of the quantum MI transitions.	
 
\begin{figure}[h!]
		\begin{center}
			\includegraphics[width=7cm,clip]{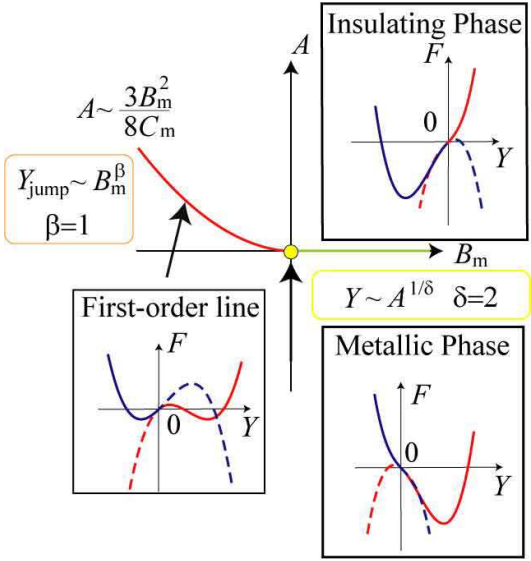}   
		\end{center}
	\caption{(color online) Phase diagram in $A$-$B_{m}$ plane.
	The first-order transition line is given 
	by the condition $A\sim 3B_{m}^2/8C_{m}$ with $B_m<0$, which separates
       the metallic phase $A<3B_{m}^2/8C_{m}$ from the insulating phase in the other side. 
       For $B_m>0$
       the transition becomes continuous and the transition line is given by 
       $A=0$ (see Eq. (\ref{ABm2})).
	Typical $Y$ dependence of the free energy  along the first-order transition line,
	in the metallic and insulating phases are also shown in the insets. 
       The definition of the free energy is
	given in Eqs. (\ref{eq:ex_free_m}) and (\ref{eq:ex_free_i}).
	The real $Y$ dependence of the free energy is represented by the solid lines, 
       where the metallic (insulating) side corresponds to
	the region $Y>0$ ($Y<0$), respectively.
	Broken lines represent the behaviors of Eqs. (\ref{eq:ex_free_m}) and (\ref{eq:ex_free_i})
       extrapolated to the other sides $Y<0$ and $Y>0$, respectively. For the derivation of the
       phase boundary, see the text.}
	\label{fig:Scheme_GL}
	\end{figure}

	Phase diagram of the MI transitions in the $A$-$B_{m}$ plane
       is shown in Fig.~\ref{fig:Scheme_GL} for the region close to $A=B_m=0$. 
	Metallic states are realized when the
	minimum is at a positive $Y_{m}$, whereas insulating states are
       stabilized when the minimum is at a negative $Y_i$.
       From Eq. (\ref{eq:MQCP_GL}), a local minimum appears at 
      \begin{equation}
		Y_{m}=\frac{-B_{m}+\sqrt{{B_{m}}^{2}-2AC_{m}}}{C_{m}}, \label{eq:M_mini}
	\end{equation}
       if $B_{m}^{2}>2AC_{m}$ and at
      \begin{equation}
		Y_{i}=\frac{-B_{i}+\sqrt{{B_{i}}^{2}-2AC_{i}}}{C_{i}}. \label{eq:I_mini}
	\end{equation} 
       if $B_{i}^{2}>2AC_{i}$ is satisfied.
       Because $F$ is bounded from below, there always exists the absolute minimum 
       either in the region $Y_i\le 0$ or $Y_m\ge 0$. The location of the absolute minimum and the 
       phase boundary shown in Fig.~\ref{fig:Scheme_GL} are clarified below 
       by classifying the parameter values into several cases. Then the critical exponents
       are derived in each region. We remind that $B_i>B_m$ and
       $C_i<0<C_m$ are always satisfied in all the relevant regions.
  
	\subsubsection{Quantum critical line}

	The continuous MI transition line (quantum critical line) is determined from the condition
	$A=0$ and $B_{m}>0$. Along this quantum critical line, the free-energy minimum
	exists at $Y=0$. For nonzero $A$ near the quantum critical line $A=0$,
       Eqs. (\ref{eq:M_mini}) and (\ref{eq:I_mini}) are reduced to 
    \begin{equation}
		Y_{m}\sim -\frac{A}{B_{m}} \label{eq:Meta_con}
	\end{equation}
    and
    \begin{equation}
		Y_{i}\sim -\frac{A}{B_{i}},  \label{eq:I_con}
	\end{equation} 
       respectively.
	Because the coefficient $A$ plays the role of the conjugate field to $Y$, 
       the critical exponent $\delta$ is 
	defined from the growth of $Y$ for small but nonzero $A$ as 
	\begin{equation}
		Y\sim A^{1/\delta}.
       \label{YAdelta}
	\end{equation}  
       Therefore, $\delta=1$ is obtained in both of the metallic and insulating sides.

	
	The susceptibility in the metallic and insulating phases are given by 
	\begin{align}
		\chi=\left(\frac{d^{2} F}{d (Y)^{2}}\right)^{-1}\sim \frac{1}{B_{m}}, \notag \\
		\chi=\left(\frac{d^{2} F}{d (Y)^{2}}\right)^{-1}\sim \frac{1}{B_{i}},
	\end{align}
	respectively. Along the quantum critical line, $B$ is of course always positive
	for the region $B_{m}>0$
	in whichever of 
    the insulating and metallic sides.
	Therefore, the susceptibility does not diverge and the critical exponent defined by  
	\begin{equation}
		\chi\sim A^{-\gamma}
	\end{equation}
	is given as
	\begin{equation}
		\gamma=0.
	\end{equation}
	
	As we mentioned in Eq. (\ref{alpha}) in Sec. \ref{sec:Introduction}, the critical exponent $\alpha$ does
	not represent the singularity of the specific heat in the
	case of the quantum phase transition.
       It rather characterizes the
	singularity of $\chi=\partial^{2} F/\partial A^{2}$, because in the present case
       around the quantum critical line, $A\propto g-g_c$ holds.
	Near the quantum critical line of the MI transition, the free energy $F$ at the minimum
       is scaled by $A$ as 
	\begin{equation}
	F\propto A^{2},
	\end{equation}
       as is shown by substituting Eqs. (\ref{eq:Meta_con}) and (\ref{eq:I_con}) into 
       Eqs. (\ref{eq:ex_free_m}) and (\ref{eq:ex_free_i}).
	From Eq. (\ref{alpha}), this indicates $\alpha=0$.
       The exponent $\beta$ is not well defined on the quantum critical line~\cite{comment_beta}, 
       because it is not possible to control the parameter $g$ to cause
       the phase transition without applying the field $A$ conjugate to the order parameter.

       The free energy form (\ref{eq:ex_free_m}) and (\ref{eq:ex_free_i}) with nonzero $A$, $B_i$
       and $B_m$ is equivalent to the
       case of the transition between metals and band insulators for noninteracting electron systems,
       if one replaces $Y$ with the carrier density $X$ of noninteracting electrons. 
       By utilizing the equivalence with this noninteracting system, in the present case of the 
       quantum critical line, we also obtain $\xi \propto X^{-1/2} \propto |g-g_c|^{-1/2}$ leading to
       $\nu=1/2$.  On the other hand, one can also calculate the dynamical susceptibility $\chi$
       of the carrier density in the noninteracting system, which yields $\chi/S \propto E_F^{-1} 
       \propto k_F^{-2} \propto \xi^{2}$. Here, $E_F$ is the Fermi energy. This leads to the dynamical exponent $z=2$.    

       Now we summarize the critical exponents on the  quantum critical line:
       $\alpha=0,\gamma=0,\delta=1,\nu=1/2$ and $z=2$.  This is common to both of the 
       insulating and metallic sides of the transition.

\subsubsection{Marginal quantum critical point}
		
\begin{figure}[h!]
		\begin{center}
			\includegraphics[width=7cm,clip]{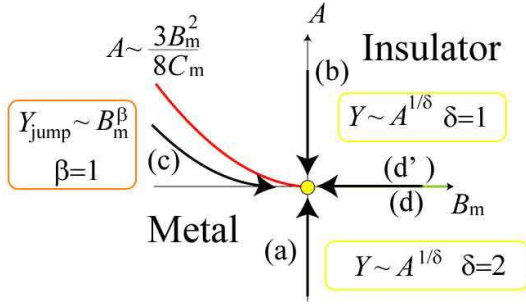}   
		\end{center}
	\caption{(color online) Four routes to approach  MQCP. Solid line in the negative $B_{m}$
	region represents the first-order transition line. The quantum critical
	line is given by the condition $A=0$ and $B_{m}>0$(see text).
	Route (a) is the way to approach MQCP by controlling ``the external field"
	$A$ from the metallic region. The singularity of the $Y$ along
	the route (a) is characterized by the critical exponent $\delta$. 
	Route (b) is the way to approach MQCP from the insulating phase.
	Route (c) is the way to approach MQCP along the first-order transition line 
       $A\sim \frac{3}{8}B_m^2$ as is derived in Eq. (\ref{ABm2}).
	Route (d) [(d$^{\prime}$)] is the way to approach MQCP along the quantum critical line
	from the metallic [insulating] phase. }
	\label{fig:ZT_CR} 
	\end{figure}

	When $B_{m}$ becomes zero, the quantum critical line terminates
	at the {\it marginal quantum critical point} (MQCP) and the first-order
	transition boundary starts. Around MQCP, we show that the exponents depend
       on the approach to the critical point either from the metallic side
       or from the insulating side.  To clarify such route dependence, 
       we introduce four routes to approach
	MQCP as is illustrated in Fig.~\ref{fig:ZT_CR}.

	From the free-energy expansion, the coefficient $A$ represents
	the conjugate field to the order parameter and 
	$B$ represents the control parameter $g$. 
    Therefore, we obtain the critical exponent $\delta$ through the
	route $(a)$ or $(b)$ in the following:  

	{\bf Route (a)}

	In the metallic side, through the route (a) in Fig.~\ref{fig:ZT_CR},
	the coefficient $A$ approaches zero by keeping $B_{m}=0$.
	From Eq. (\ref{eq:M_mini}), neglecting $B_{m}$, 
	we obtain the singularity of $Y$ as
	\begin{align}
		Y_{m}&\propto \sqrt{\frac{2|A|}{C_{m}}} \propto |A|^{\frac{1}{2}} \label{eq:Y_meta}.
	\end{align} 
	Therefore, the critical exponent $\delta$ in the metallic side is 
	\begin{equation}
		\delta=2.
	\end{equation}
       Note that this exponent is consistent with the value obtained 
       in the scaling theory~\cite{Imada_94,Imada_95,MQCP_3,MQCP_2,MQCP} 
	   as well as with that in the quantum Monte Carlo study~\cite{Furukawa_0,Furukawa_1,Furukawa_2}. 

       The dynamical exponent of the present transition
       can be calculated from the correlation functions of the order parameter
       $X$. Namely, from the equal-time structure factor $S$ for the carrier density $X$ 
       defined in Eq. (\ref{eqn:Schi2})
       and the susceptibility $\chi$ defined in Eq. (\ref{eqn:Schi3}), $z$ is given from
       Eq. (\ref{eqn:Schi1}). By calculating $S$ and $\chi$, we obtain $z=4$~\cite{z=4}.  
       Alternatively, we can derive $z$ from the following arguments: 
       Characteristic length scale of the MI
	transitions is given by the mean carrier distance $\xi$,
       which diverges at the continuous MI transition point including MQCP.
	Then, in two dimensions, carrier density $X$ is proportional to
	$\xi^{-2}$. The equal-time structure factor $S$ has the same scaling $\propto \xi^2$
       by its definition. On the other hand, from $\delta=2$, the susceptibility of 
       the carrier density $\chi$ is scaled by $X^{-1}\propto \xi^{-2}$.
	Then the dynamical critical exponent $z$ is determined
	from the condition
	\begin{equation}
		\frac{\chi}{S}\propto \xi^{z},
	\end{equation}
       as $z=4$.	 
       It also yields that $\xi$ has the singularity
	$\xi\propto |g-g_{c}|^{-\nu}$ with $\nu=1/2$.	
	The large dynamical exponent $z=4$ is consistent with
	the previous predictions obtained
	beyond the mean-field approximations~\cite{Furukawa_1, Furukawa_2, Imada_94, Imada_95}.
	
	The origin of such an unusually large $z$ is now clarified from the present study:
	The dynamical exponent normally expresses the dispersion of quantum dynamics.
	Namely, one might expect that the carrier dynamics would be described by the dispersion
	$\epsilon\propto k^4$. However, this contradicts the quasiparticle dispersion (\ref{eq:Band}),
	which is generically proportional to $k^2$.
	This puzzle is solved when we realize that the gap amplitude decreases with the
	increase of carrier number.
	Because of this reduction of the gap, when the carrier density is fixed, 
	the upper (lower) band edge as well as the Fermi level becomes lower (higher)
	than that we might expect from the rigid band picture
    obtained from the quasiparticle dispersion of insulators.
    Because of this selfconsistent reduction of the gap,
    the Fermi level does not move as quickly as $k^2$ with increasing $X$
    and is rather pinned near the original band edge even when the
    carrier number increases. This results in an effectively flat dispersion
    and actually a higher exponent with $z=4$ comes out.
    This is an intuitive picture for the filling-control transition.
    For the case of the band-width control, where the chemical potential
    is fixed rather than the filling,
    the reduction of the gap further allows more electrons and hole carriers,
    which is again interpreted by an effectively flatter dispersion.

	{\bf Route (b), (d$^{\prime}$)}

	In the insulating phase, 
	through the route (b) and (d$^{\prime}$) in Fig.~\ref{fig:ZT_CR},
	only $A$ becomes zero while $B_{i}$ remains nonzero and positive even at MQCP.
	This condition generates the same criticality as that of the
	quantum critical line. Therefore, we obtain the critical exponents 
	$\alpha=0$, $\delta=1$, $\gamma=0$ and $z=2$. 

	{\bf Route (c)}

	Next we consider the critical exponent $\beta$ defined by
	\begin{equation}
		Y_{jump}\sim |B_{m}|^{\beta},
	\end{equation}
	where $Y_{jump}$ is the jump of $Y$
	across the first-order transition line located in the region $B_m<0$.
	Near MQCP, from Eqs. (\ref{eq:M_mini}) and (\ref{eq:I_mini}),
	$Y_{jump}$ is given by
	\begin{align}
		Y_{jump}&\equiv Y_{m}-Y_{i} \\ \notag
		&\sim\frac{|B_{m}|\lambda}{C_{m}}+\frac{A}{B_i}, \label{eq:jump}
	\end{align}  
	where we define $\lambda\equiv 1+\sqrt{B_{m}^2-2AC_{m}}/|B_{m}|$.
	From this definition, it turns out that $\lambda$ is of order unity. 
	On the first-order transition line, the free energy in the metallic phase
	has the same value as that of the insulating phase, namely
	\begin{equation}
		F(Y_{m})=F(Y_{i}). \label{eq:first_F}
	\end{equation}	
	Equation (\ref{eq:first_F}) leads to
	\begin{equation}
		\frac{A\lambda |B_{m}|}{C_{m}}+\frac{B_{m}(\lambda |B_{m}|)^{2}}{2C_{m}^{2}}
		+\frac{C_{m}(\lambda |B_{m}|)^{3}}{6C_{m}^{3}}
	=-\frac{A^{2}}{2B_{i}^{2}}.\label{eq:first}
	\end{equation}
	Near MQCP, $|C_{m}|$ and $|B_{i}|$ are always nonzero and have amplitudes of the 
    order unity.
	We assume that, for small quantities $A$ and $B_m$, $A$ has the order $\epsilon_{1}$ and
	$B_{m}$ has the order $\epsilon_{2}$. From Eq. (\ref{eq:first}),
	the relation between $\epsilon_{1}$ and $\epsilon_{2}$ is given by
	\begin{equation}
		O(\epsilon_{1})O(\epsilon_{2})+O(\epsilon_{2}^{3})\sim O(\epsilon_{1}^{2}).
	\end{equation}
	From this, to satisfy Eq. (\ref{eq:first}), the relation 
	between $\epsilon_{1}$ and $\epsilon_{2}$ should be
	\begin{equation}
		\epsilon_{1}\sim\epsilon_{2}^{2}.
	\end{equation}  
       Substituting Eq. (\ref{eq:first}) into 
       Eq. (\ref{eq:M_mini}) leads to $\lambda=3/2$.
	Namely, on the first-order transition line 
	near MQCP, the relation between $A$ and $B_{m}$ should be
	\begin{equation}
		A\sim \frac{3B_{m}^{2}}{8C_m}.
       \label{ABm2}
	\end{equation}
 	Using this relation, near MQCP, the singularity of $Y_{jump}$
	is given by
	\begin{equation}
		Y_{jump}\sim\frac{3B_{m}^{2}}{8C_{m}B_{i}}+\frac{3|B_{m}|}{2C_m}\propto |B_{m}|.\label{eq:Yjump}
	\end{equation}
	Therefore, the critical exponent $\beta$ is given by
	\begin{equation}
		\beta=1.
	\end{equation}
	
	{\bf Route(d)}
	
	The route (d)  along the quantum critical line ($A=0$ and $B_{m}>0$)
	always keeps $A=0$.
	Although the susceptibility $\chi$ remains finite   
	on the quantum critical line itself ($\chi=1/B_{m}$),
	the amplitude $1/B_m$ increases toward MQCP
       and it diverges in the limit to MQCP because of $B_{m}\rightarrow 0$.  
	The singularity of the susceptibility 
	\begin{equation}
		\chi=|g-g_c|^{-\gamma}
	\end{equation}
	is given by  
	\begin{align}
		\chi&=\left(\frac{d^{2} F}{d (Y)^{2}}\right)^{-1} \\ \notag
			&\propto|B_{m}|^{-1} \propto |g-g_{c}|^{-1}.
	\end{align}
	Therefore, in the metallic phase, the susceptibility diverges at MQCP with the critical exponent
	\begin{equation}
		\gamma=1.
	\end{equation}
	
	Near MQCP, the free energy $F$ at the minimum in the metallic side is scaled
	by the control parameter $B_{m}$ as 
	\begin{equation}
	F\propto B_{m}^{3}.
	\end{equation}
       For MQCP, $B_m\propto g-g_c$ holds in contrast to $A\propto g-g_c$ for the quantum critical line.
       Then from Eq. (\ref{alpha}), we obtain the exponent $\alpha=-1$.
	We remark again that $\alpha$
	has no direct relation with the singularity of the specific heat.
       In the insulating side, as mentioned above for the route $\rm (d')$, $\alpha=0$ is obtained.

       Now the critical exponents for MQCP are summarized as 
     	\begin{equation}
        \alpha=-1, \beta=1, \gamma=1, \delta=2, \nu=1/2 \ {\rm and} \ z=4
        \label{Expsmet}
	\end{equation}
        for the metallic side
       and 
     	\begin{equation}
        \alpha=0, \beta=1, \gamma=0, \delta=1, \nu=1/2 \ {\rm and}\  z=2
        \label{Expsins}
	\end{equation}
        for the insulating side.
       The scaling laws (\ref{Widom}) and (\ref{Rushbrooke}) as well as the hyperscaling Josephson relation
       (\ref{Josephson}) associated with the hyperscaling Eq. (\ref{hyperscaling}) are all satisfied 
        in both of the two sides.


\subsection{Crossover of the critical exponent $\delta$}

	In the metallic region, $Y$ grows from zero at MQCP when $A$
	and/or $B_{m}$ deviate from zero. As a function of the distance
	from MQCP defined by $r=\sqrt{A^2+B_{m}^2}$ in the $A$-$B_{m}$ plane,
	$Y$ grows as $Y\propto r$ for $A=0$ because of Eq. (\ref{eq:Yjump})
	whereas it is scaled as $Y\propto r^{\frac{1}{2}}$ for $B_{m}=0$
	because of Eq. (\ref{eq:Y_meta}). Then a crossover between $Y\propto r$
	and $Y\propto r^{\frac{1}{2}}$ must exist. Here, we specify the location
	of this crossover boundary below.
	
\begin{figure}[h!]
		\begin{center}
			\includegraphics[width=6cm,clip]{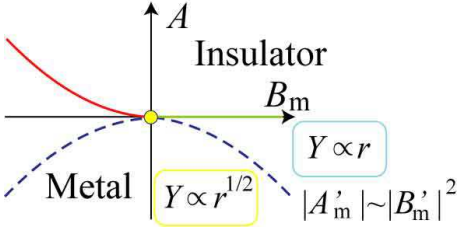}   
		\end{center}
	\caption{(color online) Crossover line between the scaling region $Y\propto r$
	 and $Y\propto r^{\frac{1}{2}}$, which  
	 is represented by the broken (blue)line defined 
	by $A^{\prime}\sim |{B_{m}^{\prime}}|^{2}$.}
	\label{fig:Crossover}
	\end{figure}

	Near MQCP, $Y_{m}$ is given as
	\begin{align}
		Y_{m}&\sim \frac{-B_{m}+\sqrt{B_{m}^2-2AC_{m}}}{C_{m}} \\
		&=-B_{m}^{\prime}+\sqrt{{B_{m}^{\prime}}^{2}+A_{m}^{\prime}},
	\label{eq:Cross}
	\end{align}	
	where we define $B_{m}^{\prime}=B_{m}/C_{m}$ and
	$A_{m}^{\prime}=-2A/C_{m}>0$.
	In the region where the condition $A_{m}^{\prime}\gg {B_{m}^{\prime}}^{2}$
	is satisfied, $Y_{m}$ behaves as
	\begin{align}
	Y_{m}\sim -B_{m}^{\prime}+\sqrt{A_{m}^{\prime}}\propto{A}^{\frac{1}{2}}. 
	\end{align}
	This indicates that MQCP exponent($\delta=2$) dominates
	in this region.
	In contrast, in the region  
	where the condition $A_{m}^{\prime}\ll {B_{m}^{\prime}}^{2}$
	is satisfied, $Y_{m}$ behaves as
	\begin{align}
	Y_{m}&\sim -B_{m}^{\prime}+B_{m}^{\prime}(1+\frac{A_{m}^{\prime}}{{B_{m}^{\prime}}^{2}}) \\
	&\sim\frac{A_{m}^{\prime}}{B_{m}^{\prime}}\propto |A|.
	\end{align}
	From this, in this region the criticality of the quantum critical line
	represented by $\delta=1$ dominates.
	In Fig.~\ref{fig:Crossover}, the crossover line is represented
	by the broken line.
 	
\subsection{Location of marginal quantum critical point}

\begin{figure}[h!]
		\begin{center}
			\includegraphics[width=7cm,clip]{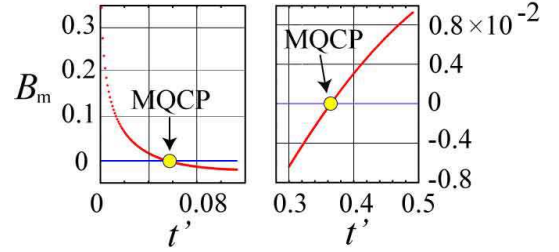}   
		\end{center}
	\caption{(color online) Coefficient $B_{m}$ on the line $A=0$($\Delta=\Delta_{c}$) near MQCP. We estimate
	the points $B_{m}=0$ as $t^{\prime}/t=0.05571$, and $0.36455$.
	In the positive $B_{m}$ region, the continuous MI transition occurs, 
	while in the negative $B_{m}$ region, the first-order
	 MI transition occurs.}
	\label{fig:All_B}
	\end{figure}

	From the free-energy form, the location of MQCP is determined by the condition 
	\begin{equation}
		A=0	\ \ {\rm and}	\ \ 	B_{m}=0. 
	\end{equation}
	The line $A=0$ is determined by 
	\begin{equation}
	\Delta=\Delta_{c}, \label{eq:C_MI}
	\end{equation}
	with the definition $\Delta_{c}=2t^{\prime}$.
	From Eq. (\ref{eq:C_MI}), the relation between $g$ and $t^{\prime}$
	for $A=0$ is given, because $\Delta$ is an implicit function of $g$ and $t^{\prime}$.
	We calculate $B_{m}$ from Eq. (\ref{eq:B_m}) with the condition
	Eq. (\ref{eq:C_MI}). We then
	find that $B_{m}$ becomes zero at two points, namely
	two MQCPs exist at zero temperature. Figure \ref{fig:All_B} shows numerical estimate of 
       $B_{m}$ near MQCP from Eq. (\ref{eq:B_m}). 
	
	From this, we estimate the location of MQCP 
	at $t^{\prime}_{c}/t=0.05571$, $g_{c}=0.616180$ and 
	$t^{\prime}_{c}/t=0.36455$, $g_{c}/t=1.38144$, respectively.
	We call MQCP at  $t^{\prime}_{c}=0.05571$, MQCP$_{1}$
	and MQCP at  $t^{\prime}_{c}=0.36455$, MQCP$_{2}$, respectively.

 \subsection{Finite temperature effect}
	In this subsection, we analyze how the free energy and critical properties are 
    modified at nonzero temperatures.
	Since the discontinuity of 
	the coefficient arising from the singularity of the density of states is 
    immediately smeared out by the Fermi distribution at $T>0$, 
	the free-energy expansion should 
	convert into the expansion of the
	conventional $\phi^{4}$ theory in the critical region.  	
	The universality class of MQCP then switches over to the Ising universality class.
	Here, we show how the expansion in Eq. (\ref{eq:MQCP_GL}) breaks
	down and the Ising critical region appears.

\begin{figure}[h!]
		\begin{center}
			\includegraphics[width=7cm]{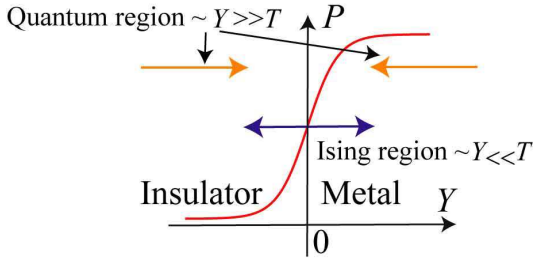}   
		\end{center}
	\caption{(color online) Schematic behavior of $P$ as function of $Y$.
	In $Y\gg T$ region $P$ behaves as the step function and the
	expansion (\ref{eq:MQCP_GL}) is valid, while in  $Y\ll T$
	region $P$ behaves as Fermi distribution function and the
	expansion (\ref{eq:MQCP_GL}) breaks down.}
	\label{fig:FT_Effect}
	\end{figure}

	The discontinuity of the coefficient $B$ and $C$
	comes from the integration of the form
	\begin{equation}
		P=\int_{-\Lambda}^{\Lambda}dE\sum_{\eta=\pm} D_{\eta}(E)H(E)G(T,E),
		\label{eq:P}
	\end{equation} 
	where $H(E)$ is a slowly varying function of $E$ in the
	range $E\sim O(T)$. Now the density of states $D_{+}(E)$
	is nonzero only in the interval
	\begin{equation}
	 -(1-\tilde{\alpha})Y<E<\Lambda,
	\end{equation}  
	whereas $D_{-}(E)$ is nonzero in the window
	\begin{equation}
	 -\Lambda<E<(1+\tilde{\alpha})Y.
	\end{equation}
	The correction factor proportional to $\tilde{\alpha}$
	arises from the chemical potential shift from the case
	of $Y=0$.
	By considering Eq. (\ref{eq:G}),    
	Eq. (\ref{eq:P}) is reduced to
	\begin{equation}
	P\cong 2H(0)\sum_{\eta}D_{\eta}(1+\tanh{\frac{(1-\eta\tilde{\alpha})Y}{2T}}),
	\label{eq:P_exp}
	\end{equation}
	if the energy dependence of the density of states $D_{\eta}(E)$
	can be ignored. Since $(1-\tilde{\alpha})/2$ and $(1+\tilde{\alpha})/2$ 
	in Eq. (\ref{eq:P_exp}) are the quantities of the order unity,
	from Eq. (\ref{eq:P}), for $Y\gg T$,  
	$P$ behaves similarly to that of the step function as shown in Fig.~\ref{fig:FT_Effect}. 	
	Therefore, for $Y\gg T$, the expansion Eq. (\ref{eq:MQCP_GL}) is valid and the MQCP criticality
	appears. We call this region  the quantum region.
	On the other hand, for $Y\ll T$, 
	$P$ behaves differently from that of the step function as shown in Fig.~\ref{fig:FT_Effect},
	and behaves as a smooth function of $Y$.
	Therefore, the expansion (\ref{eq:MQCP_GL}) breaks down and
	the conventional GLW expansion Eq.(\ref{GLFreeEnergy}) becomes justified, where
    the Ising criticality appears.
	We call this region the classical region.
	We expect that fluctuations modify the mean-field exponents for the Ising universality 
	because the upper critical dimension of the Ising universality is four.  
	We do not go into detail on this point.

\begin{figure}[h!]
		\begin{center}
			\includegraphics[width=5cm,clip]{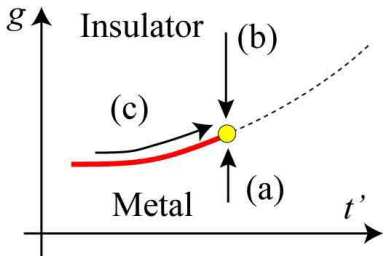}   
		\end{center}
	\caption{(color online)
	Routes to the critical point in plane of $g$ and $t^{\prime}$ at fixed 
    temperature $T$.
    Solid thick (red) line represents the first-order
	transition line and it terminates at the finite-temperature critical point
	represented by the (yellow)circle.
	Route (c) is the way to the finite-temperature critical point
	along the first-order transition line with constant $T$.
	Routes (a) and (b) are the routes to the critical point
	with fixed temperature from the metallic and insulating phases,
	respectively.  
	Broken line represents the crossover line, where $|d\Delta/dg|$ has the
	maximum value.}
	
	\label{fig:FT_Phase_2}
	\end{figure}

	Now, we clarify width of the quantum region for each route
	shown in Fig.~\ref{fig:FT_Phase_2}.
	
	{\bf Route (a)}
	
	In the quantum region,
	the singularity of $Y_{m}$ is given from Eq. (\ref{eq:Y_meta}) as 
	\begin{equation}
		Y_{m}\propto \sqrt{\frac{|A|}{C_{m}}}\sim a|g-g_{c}|^{\frac{1}{2}},	
	\end{equation}
	where $a$ is a $g$-independent constant given by
	\begin{equation}
		a=\lim_{|g-g_{c}|\rightarrow 0}\frac{1}{|g-g_{c}|^{\frac{1}{2}}}\sqrt{\frac{|A|}{C_{m}}}.\label{eq:Def_a}
	\end{equation}

	From this, the condition of quantum region $Y\gg T$
	is given by
	\begin{equation}
		|g-g_{c}|\gg \frac{T^{2}}{a^{2}}.\label{eq:Cross_a}
	\end{equation}

	{\bf Route (b)}

	In the quantum region,
	the singularity of $Y_{i}$ is given by
	\begin{equation}
		Y_{i}\propto \frac{A}{B_{i}}\sim b|g-g_{c}|,	
	\end{equation}
	where $b$ is a $g$-independent constant defined by
	\begin{equation}
		b=\lim_{|g-g_{c}|\rightarrow 0}\frac{1}{|g-g_{c}|}\frac{A}{B_{i}}.\label{eq:Def_b}
	\end{equation}
	From this, the condition of quantum region $Y\gg T$
	is given by
	\begin{equation}
		|g-g_{c}|\gg \frac{T}{b}.\label{eq:Cross_b}
	\end{equation}	

	{\bf Route (c)}

	In the quantum region, the singularity of $Y_{jump}$ is given from Eq. (\ref{eq:Yjump}) as
	\begin{equation}
		Y_{jump}\propto \frac{|B_{m}|}{C_{m}}\sim c|t^{\prime}-t^{\prime}_{c}|,	
	\end{equation}
	where a $g$-independent constant $c$ is defined by
	\begin{equation}
		c=\lim_{|t^{\prime}-t^{\prime}_{c}|\rightarrow 0}\frac{1}{|t^{\prime}-t^{\prime}_{c}|}\frac{B_{m}}{C_{m}}.\label{eq:Def_c}
	\end{equation}
	From this, the condition of quantum region $Y\gg T$
	is given by
	\begin{equation}
		|t^{\prime}-t^{\prime}_{c}|\gg \frac{T}{c}\label{eq:Cross_c}.
	\end{equation}

\section{Numerical Hartree-Fock results at $T=0$}
\label{sec:NumericalT0}
	In this section, we show  numerical results
	of MQCP obtained by numerically solving the
	Hartree-Fock self-consistent equation (\ref{eq:HF_self}) itself without relying
       on the free-energy expansion (\ref{eq:MQCP_GL}).
	We will show that numerical results are consistent
	with the free-energy expansion in the previous section.
	
 	\subsection{Phase diagram at zero temperature}
\begin{figure}[htbp]
		\begin{center}
			\includegraphics[width=7cm,clip]{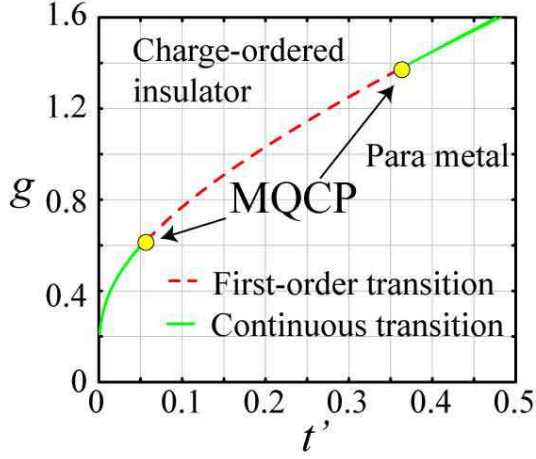}   
		\end{center}
	\caption{(color online) Phase diagram at $T=0$.
	Broken (solid) line represents the first-order (continuous)
	transition line, respectively. Near MQCP, charge-ordered metal appears.
	Detailed phase diagram near MQCP is shown in Figs. \ref{fig:ZT_Phase_S_1}(a)
	and \ref{fig:ZT_Phase_S_1}(b) }
	\label{fig:Phase_ZT_2}
	\end{figure}

	We first present the phase diagram at $T=0$ in
	Fig. \ref{fig:Phase_ZT_2}.
	We find two MQCPs in the phase diagram in agreement with the free-energy expansion.
	
\begin{figure}[h!]
		\begin{center}
			\includegraphics[width=9cm,clip]{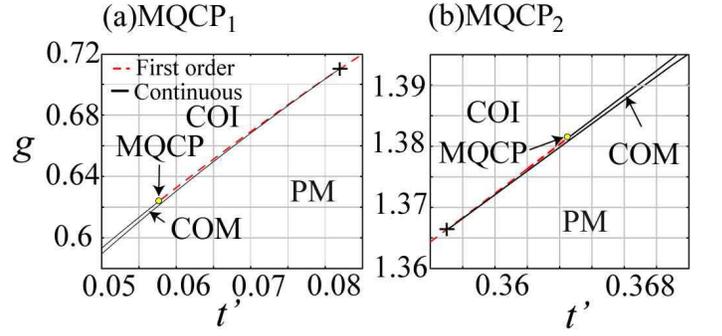}   
		\end{center}
	\caption{(color online)(a) Phase diagram near MQCP$_{1}$
	($t^{\prime}\sim 0.05571t$, $g\sim 0.6161t$).
	The continuous (first-order)  transition is shown by the
	solid (broken) line, respectively.
	MQCP is shown by the circle, where the first-order 
	transition line terminates.
	The cross point represents the triple point
	where COI, COM and PM coexist.
	(b) Phase diagram near MQCP$_{2}$
	($t^{\prime}\sim 0.3645t$, $g\sim 1.381t$). }
	\label{fig:ZT_Phase_S_1}
	\end{figure}

	At $t^{\prime}=0$, the perfect nesting occurs and the
	charge ordered phase begins from $g=0$.
	For a narrow window below the triple point ($t^{\prime}\sim 0.085t$, $g\sim 0.694t$),
	charge-ordered metal(COM)
	appears between paramagnetic-metal(PM) and charge-ordered insulator(COI) phases
    as is shown in Fig.~\ref{fig:ZT_Phase_S_1}(a) in detail. The triple point 
    represents the point where PM, COM and COI coexist.
	The order of the transitions from PM to COM is continuous,
	while the order of transition from COM to COI changes
	from the continuous to the first order with increasing $t^{\prime}/t$.
	The continuous transition line between PM and COM
	terminates at the triple point. 
	The quantum critical line between COI and COM terminates at MQCP$_{1}$
	($t^{\prime}\sim 0.05571t$, $g\sim 0.6161t$).
	
	For large $t^{\prime}$ region ($t^{\prime}>0.35t$), another triple point
	appears at $t^{\prime}\sim0.357t$, $g\sim 1.366t$  
	and COM appears between COI and PM as is shown in Fig.~\ref{fig:ZT_Phase_S_1}(b)
    in detail.
	The quantum critical line between COM and COI terminates at MQCP$_{2}$($t^{\prime}\sim 0.3645t$, $g\sim 1.381t$)
	again.      
	
\begin{figure}[h!]
		\begin{center}
			\includegraphics[width=7cm,clip]{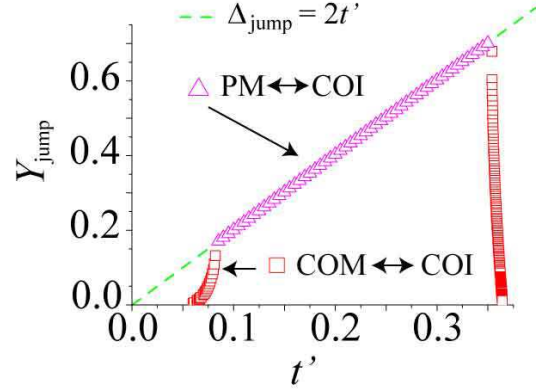}   
		\end{center}
	\caption{(color online) $Y_{jump}$ as a function of $t^{\prime}$ on the first-order MI
	transition line. For the first-order transition between PM and COI
    shown by triangles, $Y_{jump}$ is slightly larger than $\Delta_{c}=2t^{\prime}$.}
	\label{fig:Global_1}
	\end{figure}

	Figure \ref{fig:Global_1} shows the jump of $Y$($Y_{jump}$) as a function of $t^{\prime}$ 
       across the first-order MI transition line at $T=0$.
	The (red) squares and (purple) triangles show  
    $Y_{jump}$ between COM and COI and  
	between COI and PM, respectively, as a function of $t^{\prime}$.
	$Y_{jump}$ vanishes at MQCP.

	\subsection{Critical exponents of MQCP}
		We clarify the singularity of $Y$ near MQCP numerically.
		Hereafter, we only consider MQCP at $t^{\prime}_{c}=0.36455$, namely
	MQCP$_{2}$ through the route (a)$\sim$(d) in Fig.~\ref{fig:ZT_CR}. 
    Critical exponents of MQCP$_{1}$ is the
	same as  that of MQCP$_{2}$.
	We show that the numerical Hartree-Fock results are consistent with 
	the results obtained by the free-energy expansion in the previous section.   	
	
    {\bf Route (a)}
	
\begin{figure}[h!]
		\begin{center}
			\includegraphics[width=6cm,clip]{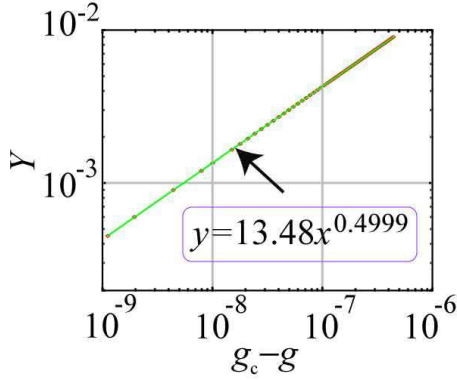}   
		\end{center}
	\caption{(color online) Log-log plot of $Y_{m}$ as function of $g_c-g$ at $t^{\prime}/t=0.36455$.
	Solid (green) line shows the result of least square fitting by assuming the function
	$y=ax^{1/\delta}$, where $y=Y_{m}$ and
	$x=(g_{c}-g)$. Results of least square fitting are given as $1/\delta=0.49996(2)$ and 
	$a=13.48(4)$.
	Critical gap $\Delta_{c}$ is obtained as $\Delta_{c}=2t^{\prime}=0.72911t$ and
	the critical interaction $g_{c}$ is estimated as $g_{c}=1.38144$.}
	\label{fig:ZT_delta_L_SC_2}
	\end{figure}

	First, we confirm that the critical exponent
	$\delta$ in the metallic region defined by
	\begin{equation}
		Y_{m}\propto |A|^{\frac{1}{\delta}} \ \  {\rm and} \ \ Y_m\sim a|g-g_{c}|^{\frac{1}{\delta}}
	\end{equation}
	is consistent with the value obtained by the free-energy expansion.
	We perform least square fitting  for $Y_{m}$ as a function of $g_{c}-g$ near MQCP.
	Figure \ref{fig:ZT_delta_L_SC_2} shows the results of the least square fitting.
	We estimate the critical exponent $\delta$ as 
	\begin{equation}
		\delta=2.0001(6).
	\end{equation}
	This is consistent with $\delta=2$ obtained by the free-energy expansion.
	We estimate the coefficient $a$ defined in Eq. (\ref{eq:Def_a}) as
	\begin{equation}
		a=13.48(4).\label{eq:coe_a}
	\end{equation}
	
	{\bf Route (b)}

	We also obtain the critical exponent $\delta$ in the insulating region.
	The definition of $\delta$ is given by
	\begin{equation}
		Y_{i}\propto |A|^{\frac{1}{\delta}} \ \ {\rm and} \ \ Y_m \sim b|g-g_{c}|^{\frac{1}{\delta}}
	\end{equation}
	The numerically obtained critical exponent $\delta$ is 
	\begin{equation}
		\delta=1.000000(4).
	\end{equation}
	The coefficient $b$ defined in Eq. (\ref{eq:Def_b}) is estimated as
	\begin{equation}
		b=0.996429(4).\label{eq:coe_b}
	\end{equation}
	Now it was confirmed that the Hartree-Fock result
	is consistent with the exponent $\delta=1$ obtained by the free-energy expansion.
	
	{\bf Route (c)}

	Next, we consider the critical exponent $\beta$ defined by 
	\begin{equation}
		Y_{jump}\propto |B_{m}|^{\beta}\propto c|t^{\prime}-t^{\prime}_{c}|^{\beta}.
	\end{equation} 
	We estimate the critical exponent $\beta$ as
	\begin{equation}
	 \beta=1.01(1).
	\end{equation}
	This critical exponent is consistent
	with $\beta=1$ obtained by the free-energy expansion.
	The coefficient $c$ defined in Eq. (\ref{eq:Def_c}) is estimated as
	\begin{equation}
		c=35.2(1).\label{eq:coe_c}
	\end{equation}
	
	\begin{figure}[hhh]
		\begin{center}
			\includegraphics[width=6cm,clip]{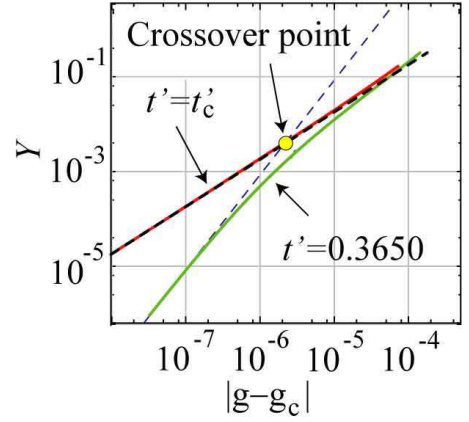}   
		\end{center}
	\caption{(color online) Crossover points of the critical exponent $\delta$ determined from
    $g-g_c$ dependence of $Y$ for  $t^{\prime}=0.3650$ near MQCP$_2$
    at $t^{\prime}=0.364555$.
	Away from the transition point, behavior of $Y$ for $t^{\prime}=0.3650$
	asymptotically  approaches that of $Y$ for $t^{\prime}=0.364555$.
	Crossover point is determined from the crossing point of $a|g-g_{c}|^{1/2}$
	and $a^{\prime}|g-g_{c}|$. The $t^{\prime}$ dependence of the crossover point
	is shown in Fig.~\ref{fig:Cross}. } 
	\label{fig:DELTA_CROSS_2}
	\end{figure}%
	\begin{figure}[hh]
		\begin{center}
			\includegraphics[width=7cm,clip]{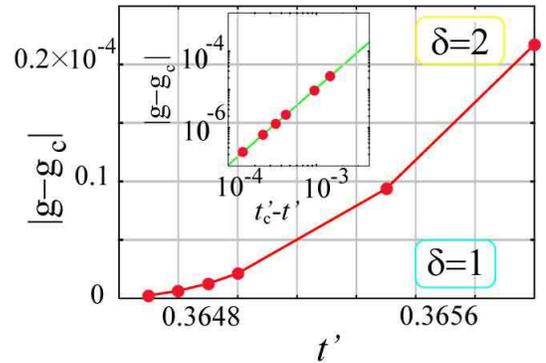}   
		\end{center}
	\caption{(color online) Crossover boundary between $\delta=2$ region
	and $\delta=1$ region in plane of $g-g_{c}$ and $t^{\prime}$.
	The inset shows log-log plot of crossover in plane of $|g-g_c|$ and $t_c^{\prime}-t^{\prime}$.
	Solid (green) line shows the result of least square fitting near MQCP
	by assuming the function $y=kx^{2}$, where  $y=|g-g_{c}|$ and
	$x=|t^{\prime}_{c}-t^{\prime}|$.}
	\label{fig:Cross}
	\end{figure}%

	\subsection{Crossover of the critical exponent $\delta$}
	The critical exponent $\delta$ has different values between the quantum critical line and 
       MQCP and it may show a crossover when the critical regions meet.  
       Here, we show numerical results for the crossover of the critical exponent
	$\delta$ at $T=0$. As already mentioned,
	at MQCP, singularity of $Y$ is well fit by the function $a|g-g_{c}|^{1/\delta}$
	with $\delta=2$ and $a=13.48$.
	Away from MQCP, $\delta$ crosses over from 2 to
	1 near the quantum critical line. In Fig.~\ref{fig:DELTA_CROSS_2}, 
       we show the least 
	square fitting for $g_c-g$ dependence of $Y$ by assuming the function 
       $a^{\prime}|g-g_{c}|^{1/\delta}$, where $\delta=1$.
	The crossover point is estimated by the crossing point of the function $a|g-g_{c}|^{1/2}$
	and $a^{\prime}|g-g_{c}|$, which is illustrated in Fig.~\ref{fig:Cross}.
	The inset in Fig.~\ref{fig:Cross} shows that the $t^{\prime}$ dependence of the crossover point
	is well fit by the function $|t^{\prime}-t^{\prime}_{2}|^{2}$.
	This is consistent with the analytical result.

\section{Numerical Hartree-Fock results at $T\ne 0$}
\label{sec:NumericalT1}

	To understand quantitatively,
	we numerically study how critical exponents of MQCP
	cross over to those of Ising critical points by finite-temperature effects.
	As shown in the previous section, the free-energy expansion predicts that crossovers of
	the critical exponents depend on the routes (a)$\sim$(c).
    
	{\bf Route (a)}
    
\begin{figure}[h!]
		\begin{center}
			\includegraphics[width=7cm,clip]{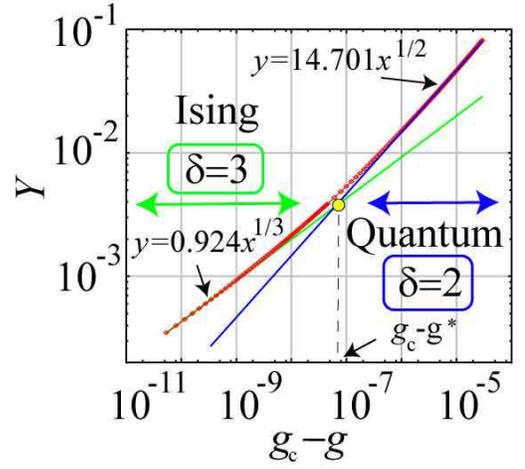}   
		\end{center}
	\caption{(color online) Log-log plot of $Y_{m}$ as function of $|g-g_c|$ near 
	the critical point at $T=0.001t$.
	Two thin solid (green and blue) lines show 
	the results of least square fitting by assuming the function
	$y=ax^{1/3}$ and $y=a^{\prime}x^{1/2}$, respectively,
	where $y=Y_{m}$ and $x=g-g_{c}$. 
	We obtain $a=0.924(1)$ and $a^{\prime}=14.701(3)$.}
	\label{fig: T1_SC_metal}
	\end{figure}

	In the metallic phase, we show the crossover of the 
	critical exponent $\delta$ near the critical point 
	for an example of $T=0.001t$ in Fig.~\ref{fig: T1_SC_metal}.
	Here, $\delta$ clearly crosses over from $2$(MQCP)
	to $3$(Ising mean-field value). We estimate the width of Ising region
	by the crossing point of two solid (green and blue) asymptotic lines.
	The value of $g$ at the crossing point is denoted as $g^{*}$.
	The Ising region is given by $W<W^*$ for
	\begin{equation}
		W\equiv |g_{c}-g|\label{eq:Def_W}
	\end{equation}  
       and $W^*\equiv |g_c-g^*|$.
       In the example in Fig.~\ref{fig: T1_SC_metal}, the crossover occurs at 
	\begin{equation}
	 W^{*}=0.6\times 10^{-7}t.
	\end{equation}
	On the other hand, the free-energy expansion in Eq. (\ref{eq:Cross_a}) indicates that the crossover is
	given by 
	\begin{equation}
		W_{FE}^{*}\sim \frac{T^{2}}{a^{2}}t\sim \left(\frac{0.001}{13.48}\right)^2t\sim 10^{-8}t,	
	\end{equation}
	where $a$ is estimated in Eq. (\ref{eq:coe_a}).
  	They are consistent each other.
	
\begin{figure}[h!]
		\begin{center}
			\includegraphics[width=9cm,clip]{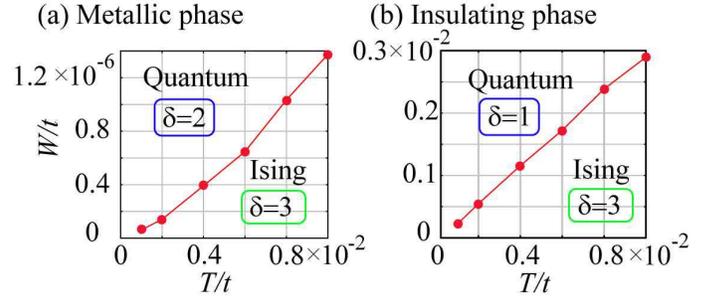}   
		\end{center}
	\caption{(color online)(a) Phase diagram of crossover boundary between 
	Ising and the quantum regions for $\delta$
	in the metallic phase plotted in plane of $W/t$ and $T/t$.
	(b) Phase diagram of crossover boundary between 
	Ising and quantum regions for $\delta$
	in the insulating phase plotted in plane of $W/t$ and $T/t$.}
	\label{fig:All_Cross_delta_m}
	\end{figure}

	For several choices of temperatures, the same analyses were performed and 
	we determined the Ising region. We show the result 
	in Fig.~\ref{fig:All_Cross_delta_m}(a).
	In the free-energy expansion, the crossover $W^{*}$
	is expected asymptotically as $W^{*}\sim T^{2}$ at low temperatures.
	The scaling of $W^{*}$ as a function of $T$ appears to show somewhat smaller 
       exponent $p$ than two for the fitting $W^{*}\sim T^{p}$. 	
       The origin of this discrepancy is likely to
	come from temperature dependence in $a$.
	Except for this discrepancy, the order of the Ising region estimated
	from the Hartree-Fock calculations is consistent
	with that of the free-energy expansion.

	{\bf Route (b)}

	In the insulating region, we show the numerically obtained 
       phase diagram for the crossover in Fig.~\ref{fig:All_Cross_delta_m}(b).
       To obtain the crossover point from $\delta=3$ for the Ising region to $\delta=1$
       for MQCP, we have used a criterion different from the metallic side in the route (a).
       The slope of $Y$ as a function of $W$ is not a monotonic function of $W$ and 
       it crossovers first from $\delta=3$ to an even smaller value, after which it increases to
       $\delta=1$.  To specify the crossover point, we have taken it as the point which shows 
       the minimum slope in the crossover region. 
	For example, at $T=0.001t$, $W^{*}$ is estimated as
	\begin{equation}
		W^{*}=2.23\times 10^{-4}t.	
	\end{equation}	
	The free-energy expansion indicates from Eq. (\ref{eq:Cross_b}) that the crossover
	point $W^{*}$ for $\delta$ is given by
	\begin{equation}
	W_{FE}^{*}\sim \frac{T}{b}\sim \frac{0.001t}{0.99}\sim 10^{-3}t,
	\end{equation}
	where $b$ is estimated in Eq. (\ref{eq:coe_b}).
	As far as the order of $W$,
	it indicates that the Hartree-Fock numerical results are consistent with
	that of the free-energy expansion. 			
	Figure \ref{fig:All_Cross_delta_m}(b) is obtained by performing 
    the same analyses for several choices of temperatures.
	The temperature dependence
	of the crossover is estimated as $W^{*}\sim T^{1.1}$.
	This is rather consistent with $W^{*} \sim T$ obtained from
	the free-energy expansion.
       It is likely that a slight discrepancy between two results comes from
	the temperature dependence of $b$.

	{\bf Route (c)}
	
	We also discuss the crossover of the critical exponent $\beta$ in the route (c).
       For example, near the critical point at $T=0.001t$, $\beta$
	crosses over from $1$(MQCP) to $1/2$(Ising mean-field) in a monotonic fashion.
       The crossover
      $W^{*}$ is estimated by the same method as the route (a) as
	\begin{equation}
	    W^{*}=0.00037\sim 10^{-4}t.
	\end{equation} 
	In the free-energy expansion,
	the crossover is estimated from Eq. (\ref{eq:Cross_c}) as
	\begin{equation}
		W_{FE}^{*}\sim \frac{T}{c}=\frac{0.001t}{32.4}\sim 10^{-4}t,
	\end{equation}
	where $c$ is estimated in Eq. (\ref{eq:coe_c}).	
	The numerical Hartree-fock result is consistent with $W_{FE}^{*}$.

\begin{figure}[h!]
		\begin{center}
			\includegraphics[width=6cm,clip]{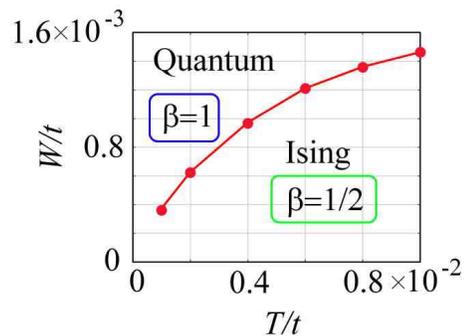}   
		\end{center}
	\caption{(color online) Crossover boundary for the critical exponent $\beta$
	plotted in plane of $W/t$ and $T/t$.}
	\label{fig:Cross_1219}
	\end{figure}

	Figure \ref{fig:Cross_1219} shows $W$ at the crossover for several choices of 
       temperatures.
	The temperature dependence
	of Ising region is estimated as $W^{*}\sim T^{0.7}$.
	which has somewhat smaller slope than $W^{*} \sim T$ expected from
	the free-energy expansion.
       Again, it is likely that the discrepancy comes from
	the temperature dependence of $c$.
       Numerical results suggest that we need to consider the overall temperature dependences
       of $a$,$b$ and $c$ on the quantitative level.
       However, we do not go further into details on this point.
       At least the existence of two critical regions governed by MQCP and Ising classes
       is now well established with crossovers, which are qualitatively consistent with the free energy
       expansion.  The numerical results show how the asymptotic scaling for the location of 
       the crossover derived in the free energy
       expansion are modified in regions distant from the critical point.

\section{Comparison with experimental results}
\label{sec:Comparison}	

In real materials, strictly speaking, 
well defined Mott critical point has been observed so far only at finite temperatures, 
as we discuss below.  There, 
Ising critical exponents should be observed in the region 
sufficiently close to the finite-temperature critical point.
However, because the Ising region is not
wide enough at low temperatures and crosses over to MQCP criticality,
quantum critical exponents may be observed in actual experiments.
Although we have to be careful in comparing our results for the very simple model with possible
artifact of the Hartree-Fock approximation with experimental results 
on more complicated systems, it may offer an insight into the relevance of the quantum criticality.
Here, we compare our results with the experimental results of
$\kappa$-(ET)$_{2}$Cu[N(CN)$_{2}$]Cl reported by Kagawa, Miyagawa and Kanoda~\cite{Kanoda_MI_Nature}.
The critical exponents in the metallic phase
show a clear evidence of the crossover in our numerical results for $\delta$ and $\beta$.
At $T_{c}=0.01t$(see Fig.~\ref{fig:All_Cross_delta_m}(a)), the classical Ising
region is observed for $W<10^{-6}t$. This critical temperature is comparable to the experimental
value $T_c\sim 40 K$ with the estimated transfer $\sim 0.1$- 0.2 eV obtained from 
the Hueckel calculation~\cite{T_Mori}. 
In $\kappa$-(ET)$_{2}$Cu[N(CN)$_{2}$]Cl, 1.0MPa roughly corresponds to
$0.001t$~\cite{T_Mori, Miyagawa}, which is estimated from the pressure dependence 
together with the dependence on the anion species in the experimentally obtained phase diagram.
The smallest distance from  the critical point accessible in the experimental resolution 
is estimated as $|P-P_{c}|\sim 0.1$MPa$\sim10^{-4}t$, which is much larger than $10^{-6}t$ suggested from
the theoretical estimate above.
This indicates that, in the experiments, the distance between the true critical point and 
the available point closest to it 
is two orders of magnitude larger than the distance of the crossover boundary from the critical point.
This is consistent with the experimental result
that the singularity of the conductivity is 
well fit only by the quantum critical exponent $\delta=2$.  
Other exponents in this $\kappa$-ET compound shows
$\gamma=1$ and $\beta=1$. These are completely
consistent with the present MQCP universality. 

We have also obtained that $\delta$ in the insulating phase 
shows a crossover. The critical exponent $\delta$
changes into $\delta=3$(Ising) from $\delta=1$ (MQCP).
We estimate the Ising region as $W<W^{*}\sim 10^{-3}t$ for $T_{c}=0.01t$.
This $W^{*}$ corresponds to about $1.0$MPa by
assuming that $1.0$MPa corresponds to $0.001t$.
The Ising region in the insulating phase is
much wider than that of the metallic phase.
From this, we have more chance for observing the Ising region ($\delta=3$)
in the experiments.
The experimental result shows that  
effective charge gap $\Delta$ closes
with the singularity as $\Delta\sim (P_{c}-P)^{0.4\pm 0.1}$
in the insulating phase. This indicates that $\delta=3\sim2$.
This critical exponent may be consistent with the Ising mean-field value $\delta=3$. 
If one were able to approach the Ising critical region ($|P-P_{c}|\ll 1.0$MPa)
in the organic conductor of $\kappa$-ET compound, 
since the system is quasi-two-dimensional, 
the 2D Ising critical exponent $\delta=15$ should be obtained.
It could show a crossover from the quantum exponent $\delta=1$
governed by MQCP to the Ising mean-field value $\delta=3$
in the region closer to the critical point with an extremely 
narrow critical region of $\delta=15$.
However, since the experimental data are obtained in the parameter space not close enough to the Ising region
of the critical point,
the mean-field critical exponent may be obtained similarly to
the case of (V$_{1-x}$Cr$_{x}$)$_{2}$O$_{3}$~\cite{V2O3_S}.
Although the pressure dependence of the parameter value for the transfer
is so far not available, high critical temperature $\sim 450$K of (V$_{1-x}$Cr$_{x}$)$_{2}$O$_{3}$ 
likely enables the 
observation of the Ising criticality. 

As we mentioned in Sec.~\ref{sec:Introduction}, some of the high-$T_c$ cuprates such as 
Bi$_2$Sr$_2$CaCu$_2$O$_{8+y}$ and (La,Sr)$_2$CuO$_4$ show consistency with $\delta=2$,
whereas some of other compounds such as (Ca,Na)$_2$CuO$_2$Cl$_2$ shows a quicker shift, 
rather consistently with $\delta=1$.  This variety suggests that parameter values of 
various copper oxides correspond to a diverse range from the region close to MQCP to
that deeply in the quantum critical line. The normal state properties as well as superconducting 
transition temperatures may depend on the distance from MQCP, which is an intriguing issue
left for the future. In the case of filling-control transitions, the first-order transition 
appears as an intrinsic instability to the phase separation. The phase separation is transformed to 
the electronic inhomogeneity with finite-size domains under the constraint of the long-range Coulomb 
interaction and the charge neutrality.  The distance from MQCP directly measures the tendency for the instability toward such inhomogeneities.  The diversity of the cuprate superconductors in terms of this tendency can be classified by this distance.  

$^3$He adsorbed on a graphite surface offers a unique purely two-dimensional fermion system. In addition to
the anomalous suppression of $T_F$ that is consistent with $z=4$~\cite{Saunders} as mentioned in Sec.\ref{sec:Introduction},
recent detailed study at lower temperatures has revealed richer structure in the specific heat~\cite{Fukuyama}.
The second layer of He approaches the registered phase (namely, a
solid phase with commensurate periodicity to the first-layer solid at the density 4/7 in terms of the first layer density)
with increasing He pressure. Near but below the density at the registered phase, the lower temperature structure of the 
specific heat suggests that the large residual entropy expected from the large dynamical exponent $z=4$ is
released below 10mK.  
A strong crossover around 10$\%$ doping between low and high density regions is suggested from the transfer of the entropy release, where the release from the specific-heat peak structure at 10-100mK 
for the lower density (more than 10$\%$ doping) region is replaced with the release at lower temperatures around or less than 1mK for 
the higher density (less than 10 $\%$ doping). The release at this low-temperature region may arise from the 
superposition of the contribution from carrier motion and spin, which are originated from two different regions of Brillouin zone, respectively, as 
analyzed by the electron differentiation in momentum space with a simplified two fluid picture~\cite{MQCP_3}.
This transfer of the entropy-releasing temperature implies a drastic 
topological change of the Fermi surface in the ground state such as Lifshitz transition to the small Fermi pocket or approximate arc structure. If a clear Lifshitz transition occurs, we expect a singular dependence of thermodynamic properties on the density measured at low temperatures below 1mK. Another conceivable case is that such a singular dependence is not observable when the topological change occurs in the pseudogap region with faint spectral weight out of the "arc".  
This may be analyzed by measuring the data at temperatures lower than 1mK in more detail 
around this crossover, which is an intriguing issue left for future studies.    
If Lifshitz transition occurs, the transition to the registered phase occurs also as a topological transition with vanishing Fermi pockets to the registered phase.  
So far, it is not clear whether the transition to the registered phase is eventually realized across
the quantum critical line or near MQCP, because the lower temperature data below 0.1mK is not available.

Physics of electron differentiation tightly associated with the present unusual universality of metal-insulator (or liquid-registered phase) transitions is a common central issue of the high-$T_c$ cuprates and monolayer $^3$He. 
The overall experimental indications show the relevance of the unusual quantum criticality
around MQCP.
More detailed analysis is left for future studies.   
Studies on the crossover effect beyond the Hartree-Fock approximation 
are also left for future studies.

\section{Summary and discussion}
\label{sec:Summary}
	Quantum MI transitions between
	symmetry broken metals and insulators
	have been studied within the Hartree-Fock approximation
	using the extended Hubbard model.
	By increasing the onsite Coulomb interaction $U$ and 
	nearest-neighbor Coulomb interaction $V$, this model at half filling
	has a tendency to charge ordering or antiferromagnetic ordering
	within the mean-field level.
       Preexisting gaps generated by such spontaneous symmetry breakings allow us
       to capture the essence of electron correlation effects on metal insulator transitions. 
	In this paper, we have mainly considered the case of the charge ordering in the region $4V>U$.
       However, it is essentially the same for the antiferromagnetic ordering in the region $U>4V$
       and the present study can be easily extended, where the criticality does not change.
	
	Using the Hartree-Fock free energy,
	we have studied the criticality of MI transitions. 
	To clarify the nature of MI transitions, 
	we have performed the free-energy expansion as
	\begin{equation}
			F_{\Delta}=F(\Delta_{c})+AY+
			\frac{B}{2!}Y^{2}+\frac{C}{3!}Y^{3},\label{eq:F_5} 
	\end{equation}
	in terms of the gap amplitude measured from the MI transition
	point , $Y=\Delta_{c}-\Delta$.	
		This free-energy expansion has an unconventional feature, where the coefficients
       $B$ and $C$ have different values between $B_m$ and $C_m$ in metals and 
       $B_i$ and $C_i$ in insulators with jumps at the transition point $Y=0$.
       Although the expansion is not regular at $Y=0$, $Y$ dependence of $F$ 
       is regular in a piecewise analytic way in each metallic ( $Y>0$ ) and 
       insulating ( $Y<0$ ) region separately, which makes the expansion unique.
	Because of the jump from $C_m>0$ to $C_i<0$ at $Y=0$, the whole expansion 
       (\ref{eq:F_5}) up to the cubic order of $Y$ is bounded from below 
       and the free-energy minimum appears at a finite $Y$.
       Such jumps of the coefficients at $Y=0$ clearly violate Ginzburg-Landau-Wilson
       scheme of phase transitions and indicate that the MI transition at $T=0$ 
       is not explained by the concept of the spontaneous symmetry breaking.  
       The origin of the jumps is ascribed to the jump of the
       density of states from a nonzero value in metals to zero in insulators in
       two dimensions.  It reflects the topological nature of
       the MI transition, which is caused by the disappearance (or emergence ) 
       of the Fermi surface without a change in the symmetry. 

	From the free-energy expansion in Eq. (\ref{eq:F_5}),
	we find that both continuous and
	first-order MI transitions occur at $T=0$.
	The numerical estimates for the relevant parameter region of the extended Hubbard model 
       show $B_i>0$ and $B_i>B_m$, whereas $C_i<0<C_m$.  
       When $B_m>0$, the region $A<0$ represents
	the metallic phase, while the insulating state is stabilized for $A>0$. 
       For $B_m>0$, the MI transitions are always continuous.
	The quantum critical line with the continuous MI transition is determined
	by the condition $A=0$. When $B_m$ becomes zero,
	the quantum critical line terminates at the
	marginal quantum critical point (MQCP)[$A=0$, $B_m=0$]
	and the first-order transition line begins in the region $B_m<0$.
	Our calculated coefficients $A$ and $B_m$ specify the locations of MQCP
	at $t^{\prime}=0.05571$ and $t^{\prime}=0.36455$ for the extended Hubbard model.
	
	In Sec. 4, we have studied the MI transitions by
	solving the self-consistent equation numerically.
	The obtained phase diagram in $t^{\prime}$-$g$ plane
	at $T=0$ contains two MQCPs.
	We have confirmed that the numerical results are
	consistent with the analytical results obtained
	in Sec. 3.

	Our mean-field exponents at MQCP is essentially exact beyond
       the mean-field theory. 
	Substituting  these critical exponents to the Ginzburg criterion~\cite{Nigel},
	we obtain the upper critical dimensions $d_{c}$ as
	\begin{align}
		d_{c}=\frac{\gamma+2\beta}{\nu}-z \notag 
			=2.	
	\end{align}
	Since the two-dimensional system is at the upper critical dimension,
	the mean-field critical exponents of MQCP should be exact
	except for possible logarithmic corrections.
	
	Using the free-energy expansion in Eq. (\ref{eq:F_5}),
	we obtain the critical exponents of MQCP as
	$\alpha=-1$, $\beta=1$, $\gamma=1$, $\delta=2$, $\nu=1/2$, $\eta=0$ and the 
       dynamical exponent $z=4$ in the metallic side of the transition.
	These critical exponents show that MQCP
	belongs to an unconventional universality class different from those obtained from
       phase transitions caused by the spontaneous symmetry breaking.
    
       The discontinuities of the coefficients $B$ and $C$ exist, in the strict sense,
       only at $T=0$.
       At nonzero temperatures, the jump is immediately smeared out, which makes a crossover
       to a different universality.  We find that the first-order MI transition boundary
	extends to nonzero temperatures and terminates
	at the critical point. Around the finite-temperature critical line, the free energy
	follows the GLW scheme. The criticality is perfectly categorized by the Ising universality.
	This means that the universality class of the finite-temperature transition
	may be characterized as a symmetry breaking transition in contrast to $T=0$ transitions.

       The peculiarity of the universality at MQCP is ascribed to the fact that MQCP appears as the 
       marginal point between the conventional GLW transition at nonzero temperature with 
       emergence of the spontaneous symmetry breaking, and the topological transition at $T=0$,
       each of which are described by already known physics, while its connecting point is not.  
	This unconventional universality class of MQCP
	is consistent with recent experimental results of $\kappa$-(ET)$_{2}$Cu[N(CN)$_{2}$]Cl.  
	Although we have focused on the band-width control MI transitions,
	these critical exponents are the same as those of the filling-control
	MI transitions.
	
	We have clarified how the crossover between Ising and MQCP
	universalities appears. It has been shown that even the finite-temperature
	critical point may show MQCP region when the parameter is tuned
	away from the narrow Ising region.
	This explains the experimental observation on the $\kappa-$ET
	compound.
	 
	In this paper, we have considered only
	MI transitions in the ordered phase.
	We assume that a preexisting gap exists in the metallic phase
	because of the symmetry breaking.
	However, beyond the mean-field theory, even
	without the symmetry breaking,  
	preexisting gap opens in the metallic phase
	as in the case of Mott insulator at low dimensions.
	Therefore, the consequence of
	the present mean-field treatment may survive
	without a strict order, where $\Delta$ may be replaced
	with the correlation induced gap without long-range order.
       This is indeed corroborated by various numerical studies performed beyond the mean-field
       approximations, which consistently support the hyperscaling with $\delta=2$ and $z=4$.

       The present results show that the MI transition is governed by the topological change
       of the Fermi surface, with shrinkage (or emergence) at selected momentum points.
       This is different from other types of mean-field theories such as the dynamical mean-field
       approximations, where the MI transition is governed instead by the vanishing 
       renormalization factor $Z$.
       On the verge of the MI transition in the metallic side, it possibly means that the Fermi surface is 
       reduced to small pockets which violates the Luttinger sum rule.
       If the system undergoes a Lifshitz transition from a large to small
       Fermi surface in the metallic phase, this violation is allowed in the side of small pockets.
       The improvement of the dynamical mean-field theory to include the momentum dependence 
       indeed suggests the existence of 
       such a Lifshitz transition as we discussed in Sec. \ref{sec:Introduction}.
       The existence of the small Fermi surface without
       folding of the Brillouin zone in the absence of the translational symmetry breaking
       is a fundamental issue and should be more thoroughly studied in the future.

       The present study has revealed a mechanism of generating quantum critical point of 
       the metal-insulator transition, which has been unexplored in the literature. 
       The marginal quantum critical point 
       emerges because of the generation of the gap determined in cooperation 
       with the kinetic energy gain of metallic carriers. 
       Its combined and nonlinear effect yields pinning of the quasiparticle
       energy in contrast to the rigid band picture.
       This may be expressed by attractive effective
       interaction of carriers.
       It generates an unconventional universality.

       A remarkable finding is that MQCP is governed by the coexistence of  quantum fluctuations
       and the divergence of the density fluctuations at small wave numbers.  Such density fluctuations 
       may cause various further and deeper effects near MQCP. Its thorough understanding is 
       an intriguing issue left for future studies. Among all, revealing 
       momentum and frequency resolved density fluctuations is an extremely important
       challenging issue of experimental studies, which must reveal an anomalous and critical
       enhancement near MQCP as well as near the finite-temperature critical point if 
       accurate measurements are made. 
	   Thorough studies of superconducting mechanism arising from MQCP~\cite{MQCP_2}
	   are also extremely important issue to be explored.	
	
	\begin{acknowledgments}
	The authors are indebted to Youhei Yamaji for extensive discussions.
	The authors thank Hiroshi Fukuyama, 
	Kazushi Kanoda and Fumitaka Kagawa for illuminating discussions on their experimental results.
	One of the authors(T. M.)  thanks 
	Shinji Watanabe for stimulating discussions. 
	This work is supported by Grant-in-Aids for Scientific Research
	on Priority Areas under the grant numbers 17071003 and 16076212 from MEXT, Japan.
	A part of our computation in this work has been done using the
	facilities of the Supercomputer Center, Institute for Solid
	State Physics, University of Tokyo.
	\end{acknowledgments}
	
\appendix		
\section{}

\label{ap:A}
	In this appendix we show details of the numerical calculations.
	In the standard procedure of solving Eqs. (\ref{eq:HF_self}) and (\ref{eq:HF_density}),
	input parameters are effective interaction $g$ 
	and particle density $n$.
	In this procedure, chemical potential $\mu$ and the order parameter $m$ are
	determined by solving  Eqs. (\ref{eq:HF_self}) and (\ref{eq:HF_density}) self-consistently. 
	This procedure costs much computation time, and is not efficient
	in determining $\mu$ and $m$ precisely.
	
	Instead, we use a different procedure.
	Namely, we take $\Delta$ and $n$ as input parameters, while output
	parameters are $g$ and $\mu$.
	By following this procedure, first, in Eq. (\ref{eq:HF_density}) for fixed $\Delta$,
	we can determine the chemical potential $\mu$ to satisfy the  condition $n=1$
	independent of Eq. (\ref{eq:HF_self}).
	Then, using this $\mu$, we perform the integration of the right hand side of Eq. (\ref{eq:HF_self})
	and obtain $g$. Finally, we obtain the order parameter $m=\Delta/g$.
	Since this procedure does not require the self-consistent calculations,
	it is possible to calculate Eqs. (\ref{eq:HF_self}) and (\ref{eq:HF_density}) rapidly and precisely.
	
\begin{figure}[htp]
		\begin{center}
			\includegraphics[width=6cm,clip]{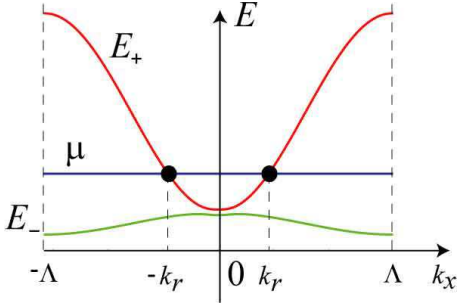}   
		\end{center}
	\caption{(color online) Schematic band structure for fixed $k_{y}$.
	Circles represent the crossing points of the upper band and the chemical
	potential, which occur at $k_r$.}
	\label{fig:Integration}
	\end{figure}

	Hereafter, we explain the details of integration of Eqs. (\ref{eq:HF_self}) and (\ref{eq:HF_density}).	
	To calculate the $k$-space integration in Eq. (\ref{eq:HF_density}), first we fix $k_y$ and
	obtain the particle density integrated over $k_{x}$, namely  $n(k_{y})$.
	It is easy to solve the equation 
	\begin{equation}
		\xi_{2}(k_{x},k_{y})\pm\sqrt{\xi_{1}(k_{x},k_{y})^2+\Delta^2}-\mu=0
		\label{eq:quadratic}
	\end{equation}
	with respect to $\cos{k_{x}}$ for fixed $k_y$ analytically since
	the equation is quadratic with respect to $\cos{k_{x}}$.
	Using the roots of the Eq. (\ref{eq:quadratic}),
	we obtain  $n(k_{y})$ exactly.
	To simplify the explanation, we only consider the case that
	the two real roots exist for upper band and their absolute values are less than 1.
	In this case, analytically we can obtain the location of the point where $E_{+}$ and $\mu$ cross,
	namely $k_{r}$ and $-k_{r}$[see Fig.~\ref{fig:Integration}].
	From this, for the upper[lower] band, the particle density for fixed $k_{y}$ is given by
	$n({k_{y}})_{+}=2k_{r}/2\pi$
	[$n({k_{y}})_{-}=2\Lambda(k_{y})/2\pi$], where $\Lambda(k_y)$ is the edge of Brillouin zone.
	The particle density for the fixed $k_{y}$ is given as $n(k_{y})=n(k_{y})_{+}+n(k_{y})_{-}$.
	In other cases, in a similar way, we can obtain $n(k_y)$ analytically.
	By integrating  $n(k_{y})$ over $k_y$, we obtain the
	whole particle density $n$. Along the $k_y$ direction, we use Newton-Cotes formula
	and take the number of points in the $k_y$ direction  up to $400000$.

	
	Next, we explain the way of integrating over $k$ in the right hand side of Eq. (\ref{eq:HF_self}).
	We fix $k_{y}$ and perform the $k_{x}$ integration first.
	Along the $k_x$ direction, we use double exponential (DE) formula,
	which is suitable for the integration of the analytic function
	in the integration period.
	It is known that DE formula is one of the optimized methods to integrate
	the analytic function with the highest accuracy using the smallest number of points.
	In the above case, integration periods $I_{1}$ and $I_{2}$
	are given as $I_{1}=[-\Lambda,-k_{r}]$ and $I_{2}=[k_{r},\Lambda]$
	(see Fig.~\ref{fig:Integration}).
	DE formula is useful in these periods, since the right hand side of Eq. (\ref{eq:HF_self}) is analytical.
	We take the number of points in these periods up to $1000$.
	Then we perform the integration over $k_{y}$ using the Newton-Cotes formula,
	and obtain the effective interaction $g$.
	We take the number of points in the $k_y$-direction up to $400000$ for the integration.

\section{}

\label{ap:B}
	In this appendix we derive
	Eqs. (\ref{eq:alpha}), (\ref{eq:D_u}) and (\ref{eq:D_l}).
	First, we obtain the density of states (DOS) at the
	bottom (top) of the upper (lower) band, respectively.
	
	The location of the bottom of the upper band is
	given by $(\pi,0)$ and $(0,\pi)$. Near $(\pi,0)$, using Eq. (\ref{eq:Band}),
	$E_{+}$ is expanded with respect to $q_{x}=k_{x}-\pi$ and
	$q_{y}=k_{y}$ as
	\begin{eqnarray}
       E_{+}&\sim& -4t^{\prime}\cos{q_{x}}\cos{q_{y}} \nonumber \\
       &&+\sqrt{4t^{2}(\cos q_{x}-\cos q_{y})^{2}+\Delta^{2}}-\mu \ \ \ \ \  \\
       &\sim& -4t^{\prime}+\Delta+2t^{\prime}q^{2}-\mu, \label{eq:E+}
	\end{eqnarray}
	where we define $q=\sqrt{q_{x}^{2}+q_{y}^{2}}$.
	From this, we obtain
	\begin{equation}
		\frac{d E_{+}}{d q}\sim 4qt^{\prime}. \label{eq:DU_1}		
	\end{equation}
	The DOS of the upper band near $(\pi,0)$ is given by
	\begin{equation}
		D_{+}=\frac{1}{4\pi t^{\prime}}.
	\end{equation}
	
	Near the MI transition, the Fermi wave number
	$q_{+}$ is determined by the condition
	\begin{equation}
		E_{+}\sim -4t^{\prime}+\Delta+2t^{\prime}q_{+}^{2}-\mu=0,
		\label{eq:F+}
	\end{equation}
	where $\mu$ is the chemical potential.
	From Eqs. (\ref{eq:E+}) and (\ref{eq:F+}), we obtain
	\begin{align}
	 	q_{+}\sim \sqrt{\frac{2\Delta_{c}-\Delta+\mu}{\Delta_{c}}},	
	\end{align}	
	where $\Delta_{c}=2t^{\prime}$.
	Using this, the electron density of the upper band,
	namely electron density $X_{+}$ including spin degeneracy
	is given by
	\begin{align}
		X_{+}&=\frac{q_{+}^{2}}{2\pi} \notag \\
			 &=\frac{1}{2\pi}\left(\frac{2\Delta_{c}-\Delta+\mu}{\Delta_{c}}\right). \label{eq:X+}
	\end{align}

	The location of the top of the lower band is
	given by $(\pi/2,\pi/2)$ and its equivalent points. 
	Near $(\pi/2,\pi/2)$, using Eq. (\ref{eq:Band}),
	$E_{-}$ is expanded with respect to $q_{x}=\pi/2-k_{x}$ and
	$q_{y}=\pi/2-k_{y}$, which leads to
	\begin{eqnarray}
		E_{-}\sim -\Delta+2q^{2}(t^{\prime}\sin{2\theta}
		-\frac{t^{2}(1+\sin{2\theta})}{\Delta_{c}})-\mu,  \ \ \ \ \ \ \ \label{eq:E-}
	\end{eqnarray}
	where we define $q=\sqrt{q_{x}^{2}+q_{y}^{2}}$.
	Then we obtain
	\begin{equation}
		\frac{d E_{-}}{d q}\sim 
		-4q(\frac{t^{2}}{\Delta_c}-(t^{\prime}-\frac{t^{2}}{\Delta_c})\sin{2\theta}). \label{eq:DL_1}		
	\end{equation}
	From Eq.~(\ref{eq:DL_1}), the DOS of the lower band near $(\pi/2,\pi/2)$ is given by
	\begin{equation}
		D_{-}=\frac{8}{(2\pi)^{2}}\int_{\frac{3\pi}{4}}^{\frac{7\pi}{4}}
		\frac{d\theta}{4(\frac{t^{2}}{\Delta_c}-(t^{\prime}-\frac{t^{2}}{\Delta_c})\sin{2\theta})} 
		=\frac{1}{2\pi \sqrt{t^{2}-{{t^{\prime}}^{2}}}}.
	\end{equation}

	Near the MI transition, the Fermi wave number $q_{-}$ is determined by the condition
	\begin{equation}
		E_{-}\sim -\Delta+2q_{-}^{2}(t^{\prime}
		\sin{2\theta}-\frac{t^{2}(1+\sin{2\theta})}{\Delta_c})-\mu=0,\label{eq:F-}
	\end{equation}
	where $\mu$ is the chemical potential.
	From Eqs. (\ref{eq:E-}) and (\ref{eq:F-}), we obtain
	\begin{align}
	 	q_{-}(\theta)\sim\sqrt{\frac{t^{\prime}(-\mu-\Delta)}
		{t^{2}-(2{t^{\prime}}^{2}-t^{2})\sin{2\theta}}}.	
	\end{align}	
	Using this, the hole density of the lower band,
	namely the hole density $X_{-}$ is given by
	\begin{align}
		X_{-}&= \frac{4}{4\pi^2}\int_{\frac{3\pi}{4}}^{\frac{7\pi}{4}}q_{-}^{2}d\theta \notag \\ 
		&=\frac{(-\mu-\Delta)}
		{2\pi \sqrt{t^2-{t^{\prime}}^2}}.\label{eq:X-}
	\end{align}

	At half filling, $X_{+}$ should be equal to $X_{-}$. From this condition, 
	recalling Eq. (\ref{eq:chemi}) and using Eqs. (\ref{eq:X+}) 
	and (\ref{eq:X-}),
	we obtain $\tilde{\alpha}$ as
	\begin{equation}
	\tilde{\alpha}=
	\frac{2t^{\prime}-\sqrt{t^{2}-{t^{\prime}}^{2}}}{2t^{\prime}+\sqrt{t^{2}-{t^{\prime}}^{2}}}.
	\end{equation}



\begin{thebibliography}{999}
\bibitem{comment}
For instance, in Refs.\cite{RMP,Imada_95}, the control parameter was taken as $\mu-\mu_c$, while here, it is taken 
as $g$, which turns out to be scaled by $(\mu-\mu_c)^{1/2}$ for the marginal quantum critical point of the 
filling-control transition in two dimensions.
This makes a trivial factor two necessary for the present $\nu=1/2$ in comparison 
with $\nu=1/4$ in Refs.\cite{Imada_95,RMP}, although $z$ is commonly 4. 
Along the quantum critical line, this factor two is not necessary, because $\mu-\mu_c$ is scaled linearly with $g$.
The scaling between $g$ and $\zeta$ will be clarified in Sec. 2. 
The exponent $\alpha=2-(d+z)\nu$ in the 
present definition is $\alpha=2-(d+z)/2$ while it is $\alpha=2-(d+z)/4$ in Refs.\cite{Imada_95,RMP}.
In any case, these transformations do not make any difference in scalings of real physical quantities.
\bibitem{z=4} Detailed derivation will be given elsewhere.
\bibitem{comment_beta} 
The critical exponent $\beta$ also characterizes
the singularity of the order parameter with respect to
the control parameter. 
Namely, we can define $\beta$ as
$X\propto a^{\beta}$, where $X$ is the order parameter and $a$
is a control parameter. In this definition, $\beta$ is well defined
even in the case of the continuous MI transitions.
In this case, since the control parameter is equivalent to the external field,
$\beta$ is equal to $\delta$ ($\beta=\delta=1$).  
In Ref.~\cite{MQCP}, this definition is chosen.
\end{thebibliography}
\end{document}